\documentclass[manuscript]{emulateapj}
\usepackage{natbib}
\usepackage{hhline}
\usepackage{graphicx}
\usepackage{xspace}
\usepackage{hyperref}
\usepackage{amsmath}
\usepackage{color}

\newcommand{\um}{\,$\mu$m\xspace}
\newcommand{\mh}{H$_{2}$\xspace}
\newcommand{\mhb}{H_{2}\xspace}
\newcommand{\chandra}{\emph{Chandra}\xspace}
\newcommand{\spitzer}{\emph{Spitzer}\xspace}
\newcommand{\ciao}{\textsc{ciao}\xspace}
\newcommand{\apec}{\textsc{apec}\xspace}
\newcommand{\behr}{\textsc{behr}\xspace}
\newcommand{\sherpa}{\textsc{sherpa}\xspace}
\newcommand{\magphys}{\textsc{magphys}\xspace}
\newcommand{\faint}{\textsc{faint}\xspace}
\newcommand{\specextract}{\textsc{specextract}\xspace}

\shorttitle{\emph{Chandra} observations of 3C\,293}
\shortauthors{Lanz et al.}

\begin{document}

\title{Jet-ISM Interaction in the Radio Galaxy 3C\,293: Jet-driven Shocks Heat ISM to Power X-ray and Molecular \mh Emission }
\author{L. Lanz\altaffilmark{1}, P. M. Ogle\altaffilmark{1}, D. Evans\altaffilmark{2}, \\
P. N. Appleton\altaffilmark{3}, P. Guillard\altaffilmark{4}, B. Emonts\altaffilmark{5}}
\altaffiltext{1}{Infrared Processing and Analysis Center, California Institute of Technology, MC100-22, Pasadena, California 91125, USA; llanz@ipac.caltech.edu}
\altaffiltext{2}{National Science Foundation, 4201 Wilson Blvd., Suite 1045, Arlington, VA 22230, USA}
\altaffiltext{3}{NASA Herschel Science Center, IPAC, California Institute of Technology, MC100-22, Pasadena, California 91125, USA}
\altaffiltext{4}{Institut d'Astrophysique Spatiale, Universit\'{e} Paris-Sud XI, 91405 Orsay Cedex, France}
\altaffiltext{5}{Centro de Astrobiolog\'{i}a (INTA-CSIC), Ctra de Torrej\'{o}n a Ajalvir, km 4, 28850, Torrej\'{o}n de Ardoz, Madrid, Spain}

\begin{abstract}

We present a 70\,ks \chandra observation of the radio galaxy 3C\,293. This galaxy belongs to the class of molecular hydrogen emission galaxies (MOHEGs) that have very luminous emission from warm molecular hydrogen. In radio galaxies, the molecular gas appears to be heated by jet-driven shocks, but exactly how this mechanism works is still poorly understood.  With \chandra, we observe X-ray emission from the jets within the host galaxy and along the 100\,kpc radio jets.  We model the X-ray spectra of the nucleus, the inner jets, and the X-ray features along the extended radio jets. Both the nucleus and the inner jets show evidence of $10^{7}$\,K shock-heated gas. The kinetic power of the jets is more than sufficient to heat the X-ray emitting gas within the host galaxy. The thermal X-ray and warm \mh luminosities of 3C\,293 are similar, indicating similar masses of X-ray hot gas and warm molecular gas. This is consistent with a picture where both derive from a multiphase, shocked interstellar medium (ISM). We find that radio-loud MOHEGs that are not brightest cluster galaxies (BCGs), like 3C\,293, typically have $L_{\mhb}/L_{X}\sim1$ and $M_{\mhb}/M_{X}\sim1$, whereas MOHEGs that are BCGs have $L_{\mhb}/L_{X}\sim0.01$ and $M_{\mhb}/M_{X}\sim0.01$.  The more massive, virialized, hot atmosphere in BCGs overwhelms any direct X-ray emission from current jet-ISM interaction. On the other hand, $L_{\mhb}/L_{X}\sim1$ in the Spiderweb BCG at z=2, which resides in an unvirialized protocluster and hosts a powerful radio source. Over time, jet-ISM interaction may contribute to the establishment of a hot atmosphere in BCGs and other massive elliptical galaxies.

\end{abstract}


\section{INTRODUCTION}

\subsection{AGN Feedback via Radio Jets}

Feedback from active galactic nuclei (AGN) is thought to play an important role in the evolution of galaxies. In numerical simulations, it has been shown to clear galaxies of gas and suppress star formation and supermassive black hole growth \citep[e.g., ][]{silk98,dimatteo05}. One type of feedback may take the form of interactions between radio jets and the interstellar medium (ISM), which may have either positive or negative effects on the star formation rate  \citep{wagner11}.  Hydrodynamical simulations of such interactions show that radio jets can create cocoons of hot X-ray emitting gas by depositing energy into the galaxy's ISM, which in turn can spread the effect of the jets to the entire host galaxy as the bubbles expand \citep{sutherland07}. In effect, radio jets can suppress star formation by driving shocks and turbulence into the ISM, thereby making the molecular gas inhospitable to forming stars, or by driving outflows that strip the galaxy of the raw materials from which stars form \citep[e.g., ][]{guillard12}.  Neutral and ionized outflows are observed in radio galaxies \citep[e.g., ][]{morganti03,morganti05b, emonts05, mahony13, holt08, lehnert11, crenshaw03, morganti13} with velocities that can exceed 1000\,km\,s$^{-1}$ and mass outflow rates of up to 60\,M$_{\odot}$\,yr$^{-1}$ for the neutral component. Additionally, a growing number of jet-driven outflows of both cold and warm molecular gas have been found recently \citep[e.g., ][]{dasyra12, dasyra14, combes13, morganti13, garcia14, fischer10, sturm11}.

\begin{deluxetable*}{llrrl}
\tabletypesize{\scriptsize}
\tablecaption{X-ray Sources\label{src}}
\tablewidth{0pt}
\tablehead{
\colhead{Name} & \colhead{Label} & \colhead{RA(J2000)} & \colhead{Dec(J2000)} & \colhead{Aperture\tablenotemark{a}} }
\startdata
UGC\,08782					&	Nucleus$+$Host 	&	13h 52m 17.81s	& +31d 26m 46.1s	& $3\farcs5$ circle; 20$''$ circle \\
CXO J135217.8 +312646			&	NC				&	13h 52m 17.79s	& +31d 26m 46.6s	& $1\farcs0\times1\farcs4$ ellipse with PA=90$^{\circ}$ \\
CXOU J135217.9 +312646		&	NE0				&	13h 52m 17.91s	& +31d 26m 46.4s	& $0.8''$ circle \\
CXOU J135212.9 +312737		&	NW1				&	13h 52m 12.9s~\,	& +31d 27m 37.0s	& $6''$ circle	\\
CXOU J135212.0 +312754		&	NW2				&	13h 52m 12.0s~\,	& +31d 27m 54.1s	& $12\farcs9\times8\farcs7$ ellipse with PA=30$^{\circ}$ 	\\
CXOU J135221.5 +312621		&	SE1				&	13h 52m 21.5s~\,	& +31d 26m 21.0s	& polygon	\\
CXOU J135224.9 +312510		&	SE2				&	13h 52m 24.9s~\,	& +31d 25m 10.0s	& $13\farcs2\times8\farcs0$ ellipse with PA=300$^{\circ}$ \\
CXOU J135214.4 +312717		&	NPS				&	13h 52m 14.4s~\,	& +31d 27m 17.5s	& $3''$ circle \\
CXOU J135225.2 +312537		&	SPS				&	13h 52m 25.2s~\,	& +31d 25m 37.3s	& $4''$ circle \\
CXOU J135218.9 +312647		&	EPS				&	13h 52m 18.9s~\,	& +31d 26m 47.6s	& $2''$ circle 
\enddata
\tablenotetext{a}{Position angles are given counterclockwise from North.}
\end{deluxetable*}

\subsection{Molecular Hydrogen Emission Galaxies}

A new class of galaxies with extremely luminous, high equivalent width \mh emission lines in the infrared (IR) was discovered using the \emph{Spitzer Space Telescope} \citep{werner04}, now referred to as molecular hydrogen emission galaxies (MOHEGs). Examples include the radio galaxy 3C\,326 \citep{ogle07}, the brightest cluster galaxy in Zw\,3146 \citep{egami06}, Stephan's Quintet \citep[NGC\,7318b; ][]{appleton06}, and the Taffy bridge \citep{peterson12}. \emph{Spitzer} IRS observations of such galaxies find mid-IR \mh luminosities of $L(\mhb)=10^{38}-10^{45}$\,erg\,s$^{-1}$, and \mh to IR luminosity ratios of $L_{\mhb}/L_{8-1000\,\mu{\rm m}}=0.001-0.1$. The MOHEG class \citep{ogle10} is defined to have L(\mh 0-0 S(0)$-$S(3))/L(PAH 7.7\um)$>0.04$, a ratio that is too large to be produced by photodissociation regions in star-forming regions \citep{guillard12}.

The molecular hydrogen responsible for this emission may be heated by three potential mechanisms.  The first is X-ray heating by active galactic nuclei (AGNs) or other X-ray sources. However, \citet{ogle10} showed that radio MOHEGs do not contain the high-luminosity, high-ionization AGN necessary to produce the observed \mh emission. The second mechanism of cosmic ray heating cannot be ruled out but requires a very high cosmic ray density \citep{ogle10}. The third mechanism of shock heating, previously seen in radio galaxies \citep[e.g.,][]{labiano13, guillard12, scharwaechter12}, is therefore deemed most likely. These shocks may be propagated due to the interaction of the radio jet and the ISM of the host galaxy. 

\subsection{3C\,293}
3C\,293 is one of brightest \mh emitters of a sample of 55 radio galaxies observed by \spitzer \citep{ogle10}, which were selected by redshift ($z<0.22$) and radio flux\footnote{$S_{\nu}{(178\,MHz)}>15$\,Jy for FR\,I and $S_{\nu}{(178\,MHz)}>16.4$\,Jy for FR\,II.}. 3C\,293 was one of the 17/55 of these galaxies to fall into the MOHEG regime with L(\mh 0-0 S(0)$-$S(3))/L(PAH 7.7\um)=0.24\footnote{Estimated within in the IRS SL/LL slits, which covers $\sim30$\% of the galactic disk.}. Its host galaxy is highly irregular and appears to be interacting with a less massive companion to the southwest. A tidal bridge connects the two galaxies, similar to the 3C\,326 system. The large scale radio morphology 3C\,293 is intermediate between Fanaroff and Riley (FR; 1974)\,I and FR\,II. \nocite{fanaroff74} Its radio luminosity $\nu L_{\nu}{\rm (178\,MHz)}=1.1\times10^{41}\,{\rm erg\,s^{-1}}$ is characteristic of an FR\,I, but it has the hot-spot morphology of an FR\,II. The bright radio core is resolved by Multi-Element Radio Linked Interferometric Network (MERLIN), revealing a double-lobed compact-symmetric source, which appears to bend out of the galactic disk. \citep{akujor96,beswick02,beswick04}. CO observations of this galaxy indicate a large \emph{cold} \mh mass of $2.2\times10^{10}$\,M$_{\odot}$ \citep{evans99, labiano13}.   Additionally, strong outflows have been detected: \citet{morganti03} and \citet{mahony13} detected a neutral H \textsc{i} outflow up to 1000\,km\,s$^{-1}$ containing $\sim10^7\,M_{\odot}$, and \citet{emonts05} detected a ionized outflow with $\sim10^5\,M_{\odot}$. \\

In this paper, we present a \emph{Chandra} observation of 3C\,293 and study the impact of jet feedback in this galaxy. Section 2 presents the observations and data reduction. We describe the X-ray morphology in Section 3 and present the X-ray spectra of various components in Section 4. In Section 5, we discuss the effect of the jet on the interstellar medium of 3C\,293 and the peculiar X-ray features observed along the large scale jets. Distance-dependent quantities in this paper are calculated using cosmological parameters $H_{0}=70\,{\rm km\,s^{-1}\,Mpc^{-1}}$, $\Omega_{M}=0.3$, and $\Omega_{\Lambda}=0.7$. With a redshift of 0.045 \citep{devaucouleurs91}, these assumptions yield a distance of 199.3\,Mpc for 3C\,293 at which 1$''$ corresponds to 0.97 kpc.

\begin{figure*}
\centerline{\includegraphics[width=\linewidth]{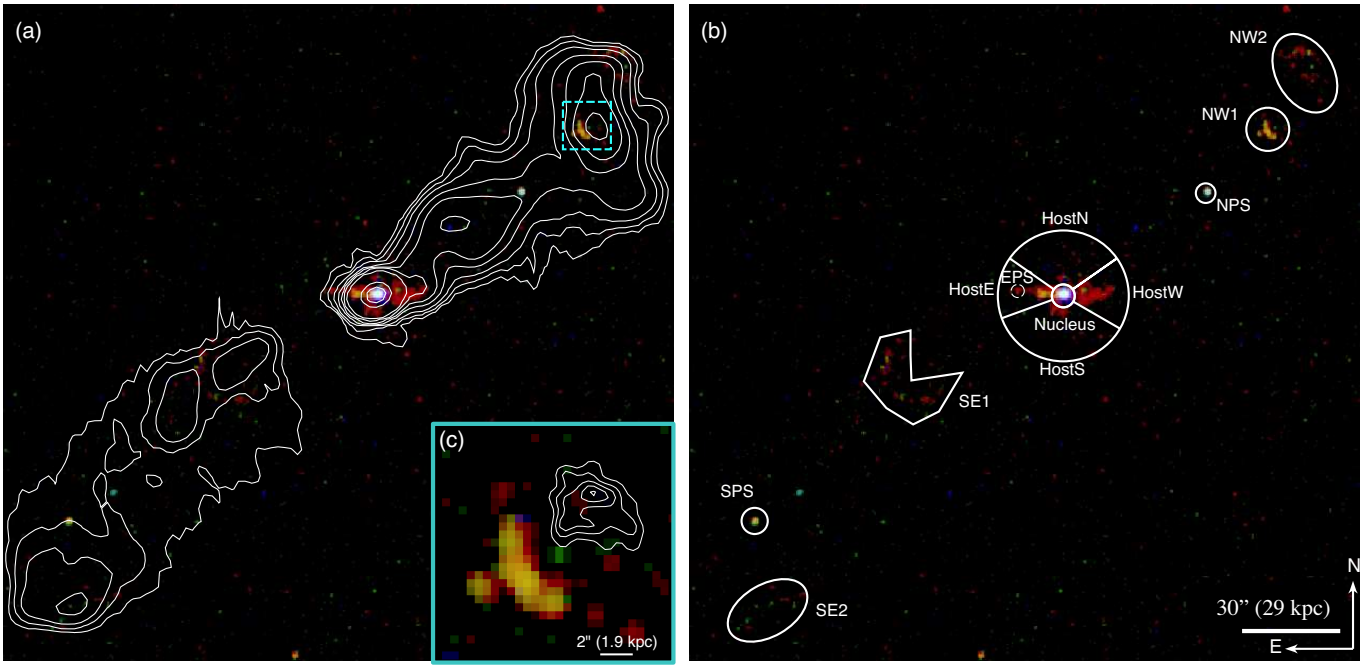}}
\caption{\emph{Chandra} image of 3C\,293 showing soft (0.5$-$1.5\,keV) emission in red, medium (1.5$-$3.0\,keV) in green, and hard (3.0$-$8.0\,keV) 
in blue, overlaid with the 1.4\,GHz contours (a; contours levels at 0.0005, 0.001, 0.002, 0.005, 0.01, 0.02, 0.05, 1.0, and 2.0\,Jy/beam) and spectral extraction apertures (b). Two jets emerge from the nucleus in the East-West direction, dominated by soft emission. Several extended features are present along the extended jets, primarily in the soft and medium bands, as well as two point sources. SE1 lies in a region of diminished radio emission compared to the jet on either side of it. NW2 and SE2 exist on the edge of the radio jets. Panel (c) shows contours of a higher resolution 1.4\,GHz image (3.25, 4.5, 5.75, 7, and 8.25\,mJy/beam) showing the location of the NW radio hotspot relative to NW1. }
\label{chandra_aps}
\end{figure*}

\section{OBSERVATIONS AND DATA REDUCTION}
\subsection{X-ray Spectra}
3C\,293 was observed for 67.8\,ks on 2010 November 16 (ObsID 12712, PI P. Ogle) with the back-illuminated CCD chip, S3, of the \chandra Advanced CCD Imaging Spectrometer \citep[ACIS;][]{weisskopf00} in \faint data mode. We reprocessed the observation using \ciao version 4.5 to create  a new level 2 events file, following the software threads from the \chandra X-ray Center (CXC)\footnote{\url{http://cxc.harvard.edu/ciao}}. We measured the counts in each aperture shown in Figures \ref{chandra_aps}b and \ref{chandra_vla_nuc} and listed in Table \ref{src}. X-ray spectra were extracted using the \specextract task in the 0.3$-$8.0\,keV energy range from ten regions of interest with at least $\sim$100 net counts. We grouped the data for each region to have a minimum of 10 counts per bin and compared the results to the background emission, measured in large source-less regions on the same chip, to determine the energy range over which the source had significant counts. We filtered the data based on that energy range and then grouped it to a minimum of 20 counts per bin prior to modeling the spectra (as described in Section 4). Counts and the luminosities derived from fitting these spectra are listed in Table \ref{cnts_flux}. 

\subsection{\emph{Spitzer} Imaging}

We retrieved Infrared Array Camera \citep[IRAC; ][]{fazio04} observations of 3C\,293 taken by the \emph{Spitzer Space Telescope} \citep{werner04} on 2005 June 11 (PID 3418, PI M. Birkinshaw) at 3.6\um, 4.5\um, 5.8\um, and 8.0\um consisting of 24$\times$30\,s frames, covering the galaxy and the nearby field. The Multiband Imaging Photometer \citep[MIPS; ][]{rieke04} observations were also obtained from the Spitzer Heritage Archive. MIPS observed 3C\,293 in the 24\um band on 2005 June 28 (PID 82, PI G. Rieke; 56$\times$2.62\,s frames), 2007 July 19 (PID 40053, PI G. Rieke; 56$\times$2.62\,s frames), and 2009 February 9 (PID 50099; 28$\times$2.62\,s frames). As part of the 2005 observation, MIPS also imaged 3C\,293 at 70\um (28$\times$10.49\,s frames) and 160\um (68$\times$10.49\,s frames). Mosaics were created using Mosaicker and Point source Extractor package \citep[MOPEX; ][]{makovoz05} for all seven bands with $0\farcs6$ pixels in the IRAC bands and $2\farcs45$ pixels in the MIPS bands.

\subsection{\emph{Spitzer} Spectral Imaging}

3C\,293 was observed with the Infrared Spectrograph \citep[IRS; ][]{houck04} on \spitzer in the mapping mode in the SL1,2 and LL1,2 modules (PID 20719, PI S. Baum; 2006 January 18). The 14\,s observations were stepped perpendicular to the slit in each module. Spectral cubes were constructed for each module using the IDL program \textsc{cubism} \citep{smith07}, using the off-pointed slit for background subtraction and removing global bad pixels. We extracted slices of the spectral cubes at the wavelengths of the rotational \mh lines (i.e., 17.03\um 0-0 S(1), 9.66\um 0-0 S(3), 6.91\um 0-0 S(5), and 5.55\um 0-0 S(7)). IRS also observed 3C\,293 in staring mode on 2007 June 19 \citep[PID 30877, PI A. Evans; previously published in][]{guillard12}. We measured the spatial extent of the \mh lines and the [NeII] line.

\subsection{Ancillary Imaging}

Ancillary images were collected from the archives of the \emph{Galaxy Evolution Explorer} \citep[GALEX in far-UV and near-UV; ][]{martin05}, Sloan Digital Sky Survey \citep[SDSS DR10 in \emph{ugriz}; ][]{ahn14} and Two Micron All Sky Survey \citep[2MASS in J, H, and Ks; ][]{Skrutskie06}. We extracted photometry in a common 30$''$ aperture centered on the X-ray nucleus in each band, as well as in the Spitzer images described above. We also retrieved \emph{Infrared Astronomical Satellite} (IRAS) photometry \citep{golombek88} from the NASA Extragalactic Database (NED)\footnote{\url{http://ned.ipac.caltech.edu}}.  

We observed 3C\,293 with the 2.4\,m Hiltner telescope of the Michigan-Dartmouth-MIT (MDM) observatory on 2007 March 14 in the B filter. Observations were done by integrating 60 min at an airmass of $1.0-1.1$ under seeing conditions of $\sim$1.2 arcsec. The observations and data reduction were performed in the same way as those described in \citet{emonts10}. 

We used archival observations taken by the Very Large Array (VLA) to make images of the source at 1.4\,GHz and 8.4\,GHz. The 1.4\,GHz image of the large scale radio jets was taken on 1999 November 18 in L band in B configuration for 1.3\,hrs \citep[PID GP022,][]{beswick04}. The 8.4\,GHz image of the core region was observed on 1995 July 27 in X band in A configuration for 0.11\,hrs (PID AK403). We also used a 1.4\,GHz VLA image taken in A configuration on 2000 November 25 (PID AM670, PI R. Morganti), whose observation and data reduction were described in \citet{emonts06} and which shows the structure of the northwest hotspot in greater detail.

\begin{deluxetable*}{lllrr}
\tabletypesize{\scriptsize}
\tablecaption{X-ray Counts and Luminosities\label{cnts_flux}}
\tablewidth{0pt}
\tablehead{
\colhead{Region} 	& \colhead{Net Counts} 	& \colhead{Hardness} & \colhead{Flux (10$^{-14}$\,erg\,cm$^{-2}$\,s$^{-1}$)\tablenotemark{b}} & \colhead{Lumin. (10$^{40}$\,erg\,s$^{-1}$)\tablenotemark{c}}  \\
\colhead{}			& \colhead{0.4$-$8.0\,keV}	& \colhead{Ratio\tablenotemark{a}} & \colhead{0.4$-$8.0\,keV}  & \colhead{0.4$-$8.0\,keV}  }
\startdata
Nucleus			& 2130$\pm$50		&  ~0.75$^{+0.01}_{-0.02}$	& $76.6^{+~\,6.9}_{-13.4}$~\,	& $364^{+33}_{-64}$~~~~~~ 			\\ %3C293_flux.xslx	
NC\tablenotemark{d} & 1810$\pm$40		&  ~0.87$^{+0.01}_{-0.01}$	& $87.6^{+~\,5.2}_{-11.8}$~\, 	& $416^{+25}_{-56}$~~~~~~ 			\\ %3C293_host_chandra.xlsx
NE0\tablenotemark{d} & ~\,194$\pm$14		&  ~0.17$^{+0.07}_{-0.07}$	& $7.22^{+1.46}_{-1.24}$		& $34.3^{+6.9}_{-5.9}$~~~			\\%3C293_host_chandra.xlsx
Host: East  		& ~\,160$\pm$14		& -0.75$^{+0.07}_{-0.07}$		& $1.38^{+0.61}_{-0.29}$	   	& $6.54^{+2.89}_{-1.37}$	     			\\%3C293_host_chandra.xlsx
Host: West 		& ~\,184$\pm$15		& -0.81$^{+0.07}_{-0.06}$		& $1.08^{+0.12}_{-0.15}$	   	& $5.15^{+0.58}_{-0.69}$	     			\\%3C293_host_chandra.xlsx
Host: North 		& ~~~87$\pm$12		& -0.32$^{+0.13}_{-0.16}$		& $0.59^{+0.24}_{-0.15}$	   	& $2.81^{+1.13}_{-0.69}$	     			\\%3C293_host_chandra.xlsx
Host: South 		& ~~~99$\pm$13		& -0.32$^{+0.15}_{-0.13}$		& $0.78^{+0.24}_{-0.17}$	   	& $3.73^{+1.13}_{-0.74}$	     			\\%3C293_host_chandra.xlsx
NW1				& ~\,119$\pm$12		& -0.58$^{+0.09}_{-0.09}$		& $1.20^{+0.28}_{-0.25}$		& $5.70^{+1.34}_{-1.18}$			\\%3C293_host_chandra.xlsx
NW2				& ~~~98$\pm$12		& -0.68$^{+0.15}_{-0.14}$		& $0.87^{+0.19}_{-0.15}$		& $4.16^{+0.88}_{-0.71}$			\\ %3C293_flux.xslx
SE1				& ~\,109$\pm$13		& -0.63$^{+0.16}_{-0.12}$		& $1.15^{+0.40}_{-0.23}$		& $5.49^{+1.91}_{-1.07}$			\\ %3C293_flux.xslx
SE2\tablenotemark{e} & ~~~47$\pm$10 		& -0.76$^{+0.07}_{-0.24}$		& $0.71^{+0.16}_{-0.15}$		& $3.37^{+0.76}_{-0.73}$			\\ %3C293_flux.xslx
NPS\tablenotemark{e} & ~~~68$\pm$~\,8 	& -0.08$^{+0.13}_{-0.12}$		& $1.42^{+0.28}_{-0.30}$		& $6.75^{+1.34}_{-1.43}$			\\ %3C293_flux.xslx
SPS\tablenotemark{e} & ~~~43$\pm$~\,7		& -0.41$^{+0.15}_{-0.17}$		& $0.56^{+0.11}_{-0.11}$		& $2.67^{+0.53}_{-0.52}$			\\ %3C293_flux.xslx
EPS\tablenotemark{f} & ~~~13$\pm$~\,4 		& -0.9					&  ...~~~~~~~~				&  ...~~~~~~~~					
\enddata
\tablenotetext{a}{Hardness ratio=(H-S)/(H+S), where H=counts in the 2$-$8\,keV range and S=counts in the 0.4$-$2\,keV range, calculated using the \behr method \citep{park06}.}
\tablenotetext{b}{Observed fluxes, uncorrected for absorption, in the aperture derived from the best-fit model in Table \ref{fits}. }
\tablenotetext{c}{Total luminosities, uncorrected for absorption, in the aperture calculated assuming a distance of 199.3\,Mpc and derived from the best-fit model in Table \ref{fits}. }
\tablenotetext{d}{NC and NE0 are contained within Nucleus, as shown in Figure \ref{chandra_vla_nuc}, but have different HR, so we also extract their spectra separately. These regions are coincident with the C and E0 regions, defined by \citet{emonts05}. } 
\tablenotetext{e}{Total luminosities for these regions were calculated assuming a power law with fixed MW absorption. } 
\tablenotetext{f}{We include EPS due to its HST counterpart, but it is only detected in the soft band, having 1 total count in the hard band.}
\end{deluxetable*}

\section{X-ray Morphology} 

Figure \ref{chandra_aps}b shows a three-color (soft, medium, and hard) X-ray image taken by \emph{Chandra} overlaid with the apertures on interesting features from which we extract integrated counts (see Table \ref{cnts_flux}) and spectra. These features fall into four types: central region (nucleus), extended emission in the host galaxy (hostN, hostS, hostE, and hostW), large scale jet-related features (NW1, NW2, SE1, and SE2), and point sources (SPS, NPS, and EPS) along the jets. Figures \ref{chandra_aps}a, \ref{chandra_aps}c, and \ref{chandra_vla_nuc} show the relation of the X-ray and radio emission. 

\subsection{Nuclear Region}

\begin{figure}
\centerline{\includegraphics[width=\linewidth]{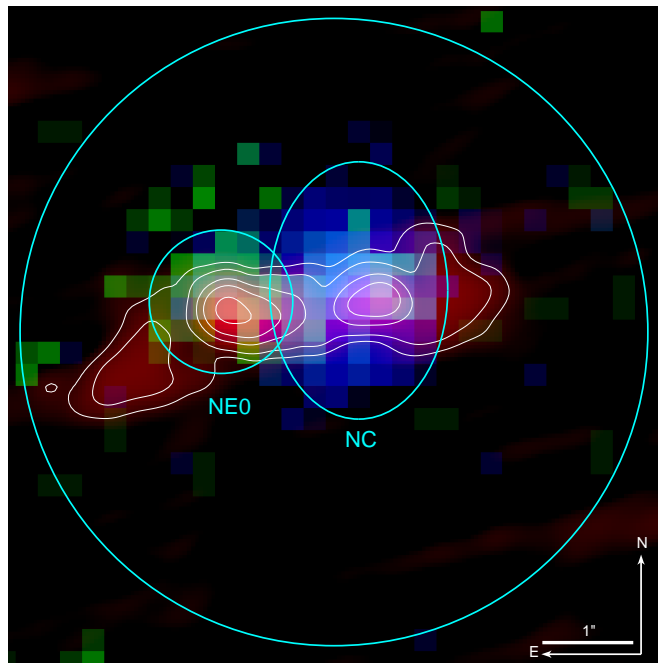}}
\caption{\emph{Chandra} (green: 0.5$-$1.5\,keV; blue: 1.5$-$8.0\,keV) and VLA (red and white contours: 8.4\,GHz) image of the nuclear region overlaid with the nuclear extraction aperture. To the east of the hard X-ray point source (NC) there is a softer component (NE0), which is approximately coincident with a knot in the very small radio jet, in turn the expected launch site of the ionized outflow.}
\label{chandra_vla_nuc}
\end{figure}

3C\,293 has a strong X-ray core (detected with $\sim2100$ counts). Figure \ref{chandra_vla_nuc} shows the nuclear region in detail with the X-ray data binned into 0$\farcs$25 pixels. High resolution radio observations \citep[e.g., ][]{beswick02, beswick04, akujor96} show a small scale radio (8.4\,GHz) jet. To the east of the nucleus, there is a softer X-ray feature coincident with a knot of radio emission, which is also the launch site of the ionized outflow \citep{emonts05}. The radio jet appears to bend $30^{\circ}$ to the southeast after about 1.7\,kpc, continuing for another $1.5$\,kpc. In contrast, the western jet does not show a bend, is only $1.4$\,kpc, and does not have a similar X-ray counterpart. The neutral outflow is associated with this western jet \citep{mahony13}.  Observations of the radio jet by \citet{akujor96} detected faint diffuse emission out to $\sim4$\,kpc, which they suggest could belong to an outburst older than the one responsible for the $<2$\,kpc jets. 

\begin{figure*}
\centerline{\includegraphics[width=\linewidth]{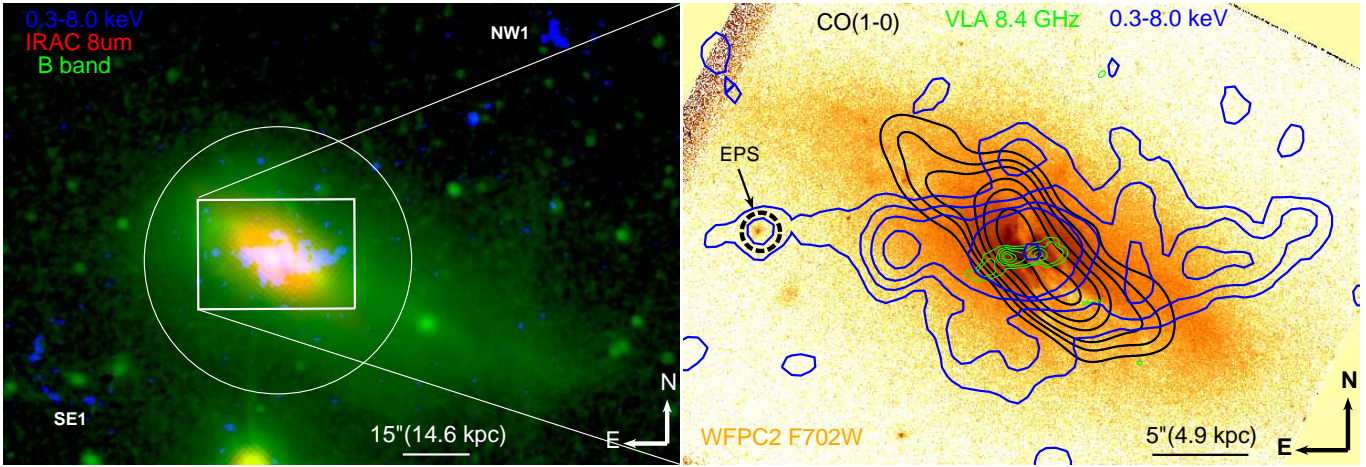}}
\caption{Multiwavelength image showing that the X-ray emission (blue) extends through most of the host galaxy's optical emission (green; B-band) and past the bulk of the MIR (red; IRAC 8.0\um). The features along the large jets (e.g., NW1 and SE1) are beyond the host galaxy and do not fall within the tidal debris extending to the southwest. A clear dust lane is visible over the nucleus, as is made even clearer in the right panel where X-ray (blue), radio (green), and CO(1-0) \citep[black; ][]{labiano13} contours are overlaid on a WFPC2 F702W \citep{floyd06} image of 3C\,293. The point source at the end of the East Jet (EPS) has a counterpart in this Hubble image. 
}
\label{spitzer}
\end{figure*}

\subsection{Host Galaxy}

Two X-ray jets (visible in Figure \ref{chandra_aps}) extend 13\,kpc to the east and west from the nucleus, similar to the radius of the galactic disk, but at a different position angle. Their emission is dominated by soft X-rays, having hardness ratios of -0.75 and -0.81, respectively. The eastern jet is narrower, and its emission is concentrated near the nucleus. In contrast, the western jet is more diffuse in both radius and angle.

Figure \ref{spitzer} shows the relation of optical and mid-IR emission to the small scale X-ray jets. There is strong IR emission in the optically-obscured dust lane, which crosses from northeast to southwest roughly in front of the nucleus. The molecular gas disk detected in CO by \citet{labiano13} has a similar morphology to the 8\um emission and is likewise most dense over the nucleus. 

We also detect X-ray emission from the host galaxy in regions perpendicular to the jets (hostN and hostS). With a hardness ratio of $-0.32$, this emission is harder than the emission in the jet regions.

\subsection{Large Scale Radio Jets}

Figure \ref{chandra_aps} shows four extended regions with significant X-ray counts along the large scale radio jet. At the end of the northwest jet's radio emission in Figure \ref{chandra_aps}a, there is a cap of primarily soft (hardness ratio of -0.68) X-ray emission, which we call  NW2. The radio emission of the northwest jet consists of two brighter regions separated by dimmer region; the region further from the nucleus spreads perpendicularly to the jet axis. On the southeast edge of this radio feature, there is a more concentrated region of X-ray emission, and it has slightly harder emission (hardness ratio of -0.58). We call this region NW1.

Along the southeast jet, we also find two extended regions with significant X-ray counts. This jet is overall dimmer in the radio bands than its counterpart. In a dimmer region between two radio knots, there is an arc of soft X-ray emission, which we refer to as SE1. At the end of the southeast radio jet, there is a brighter region (SE2) which also contains significant, predominantly soft, X-ray counts, but an insufficient number to define a spectrum for fitting. 

\subsection{Point Sources}

There are two point sources (NPS and SPS) along the large radio jets detected with significant X-ray counts, one on each side of the radio jet.  Neither contains sufficient counts to define a spectrum for fitting. SPS is likely associated with the nearby ($0\farcs44$) \spitzer source SSTSL2\footnote{Spitzer Space Telescope Source List (IRSA)} J135225.18+312537.3, whose IRAC colors are consistent with those of AGN as defined by \citet{lacy04}, suggesting SPS is a background source. Similarly, NPS may be associated with SSTSL2 J135214.38+312717.5 ($0\farcs59$ separation), which is detected in the three shortest IRAC bands. While the upper limit on the IRAC 8\um flux does not allow us to conclusively determine that it has IR colors consistent of an AGN, its 5.8\um-3.6\um color is consistent with a background AGN. The end of the inner eastern jet also has a, potentially background, point source or a knot (EPS), which has a counterpart in the HST image \citep[Figure \ref{spitzer}; ][]{floyd06}. With only 10 net counts, we can investigate this feature no further with the current observation, but we note that 13/14 total counts are in the soft band ($0.4-2.0$\,keV).

\section{X-RAY SPECTROSCOPY}

X-ray spectra were fit using the \sherpa packages of \ciao using the Levenberg-Marquardt optimization method \citep{more78}, where sufficient counts exist. A foreground absorption due to the Milky Way's ISM of $N_{H}=1.27\times10^{20}\,{\rm cm^{-2}}$ \citep{kalberla05}\footnote{\url{http://heasarc.nasa.gov/cgi-bin/Tools/w3nh/w3nh.pl}} is assumed in all fits. Where necessary, metal abundances are fixed to solar.  

\subsection{Nucleus}

\begin{figure}
\centerline{\includegraphics[width=\linewidth]{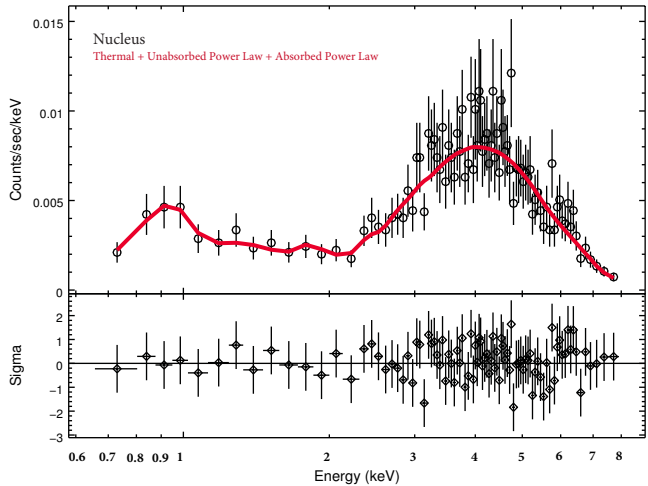}}
\caption{Spectrum of the nucleus and best-fitting model (absorbed power law + unabsorbed power law + \apec thermal model), with the residuals in the lower panel. }
\label{nuc_spec}
\end{figure}

We extracted the spectrum of the nuclear region in three apertures (see Fig. \ref{chandra_vla_nuc}). First, we examine the total nuclear emission measured in a 3$\farcs$5 aperture centered on the coordinates (13h52m17.81s, +31d26m46.1s). Figure \ref{nuc_spec} shows the resulting spectrum binned to have a minimum of 20 counts per bin. We fit the spectrum in the 0.4$-$8.0\,keV range, and found that it is best fit with a combination of thermal and power law models. An absorbed power law alone is rejected with a probability of 2.3$\times10^{-6}$. 

The best model is the sum of an absorbed power law, an unabsorbed power law, and a thermal component. The thermal component of the total nuclear spectrum was modeled with an \apec \citep{smith01} model whose metallicity was fixed at solar.  It is best fit with a temperature of 1.0\,keV. The absorbed power law component is associated with the buried AGN, while the  unabsorbed power law and thermal emission are likely associated with the off-nuclear source and diffuse emission visible in Figure \ref{chandra_vla_nuc}. We tested this by extracting and fitting two smaller regions (NC and NE0) centered on the hard emission and the off-nuclear softer emission respectively. We find that the spectrum of NC is well-fit with an absorbed power law and that the spectrum of NE0 is well fit with an unabsorbed power-law. The absorbed power law models from the total and NC spectra are broadly consistent with an absorbing column of N$_{H}=(6-9)\times10^{22}{\rm cm^{-2}}$. Similarly, the spectrum of NE0 is consistent with the unabsorbed power-law component of the total nuclear spectrum, with spectral index of 0.7.

\subsection{Host}

\begin{figure}
\centerline{\includegraphics[width=\linewidth]{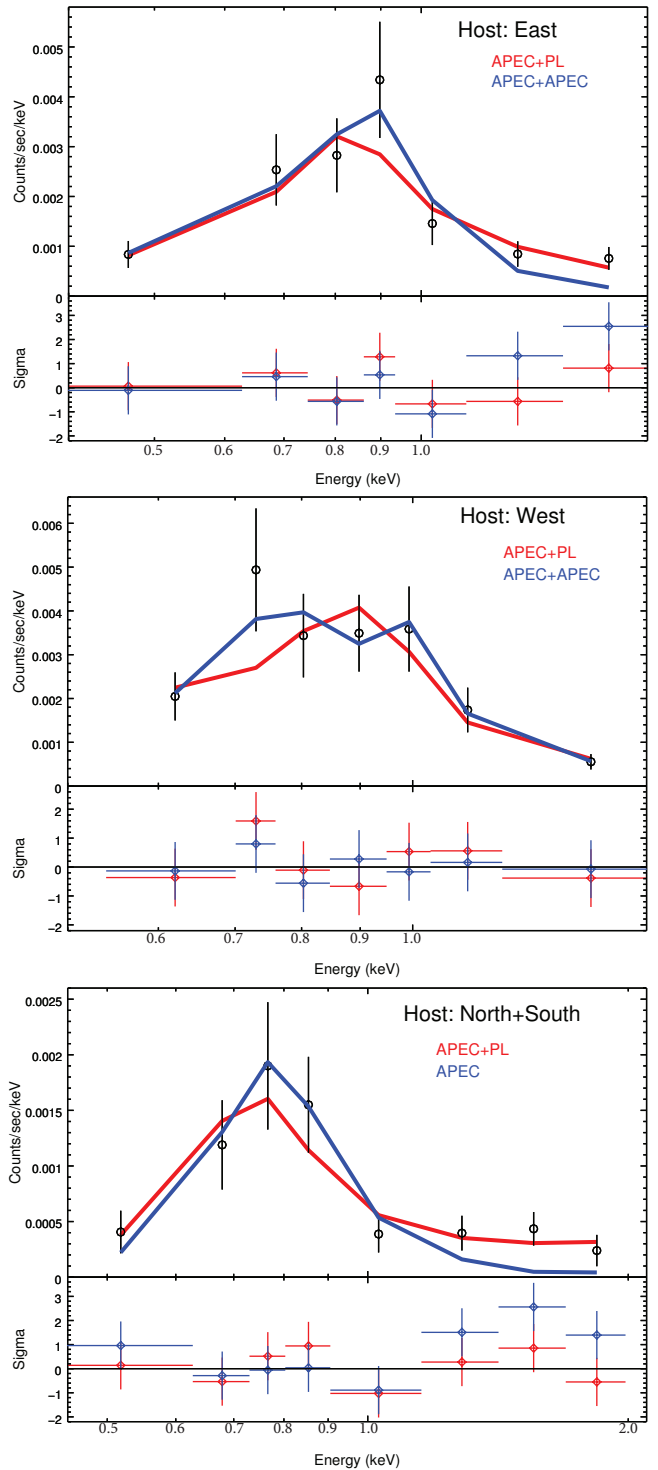}}
\caption{Spectrum of the host regions: east (top), west (middle), and the sum of north and south (bottom); and best-fitting model (power law + \apec thermal model (red) or single or sum of \apec models (blue)), with the residuals in the lower panel. }
\label{jet_spec}
\end{figure}

The spectra of the host were extracted in the four azimuthal segments of a 20$''$ circle centered on, but excluding the nucleus, as shown in Figure \ref{chandra_aps}. We combined the hostN and hostS region for greater spectral resolution on the host contents in regions not containing the jets. Figure \ref{jet_spec} shows the resulting spectrum as well as the hostE and hostW spectra. Above 2\,keV, each region only contains counts consistent with background emission so we only fit the spectrum from 0.4$-$2\,keV. We fit each spectrum with (1) a power law alone, (2) a thermal \apec model alone, (3) the sum of a power law and a thermal \apec model, and (4) the sum of two thermal \apec models, with a foreground absorber due to the Milky Way ISM. The best fit parameters are given in Table \ref{fits}. The metallicity of the \apec model is fixed at solar.

HostE is best modeled as a combination of thermal and power law emission with $\Gamma=2.1$ and a temperature $kT=0.7$\,keV (Figure \ref{jet_spec} top). Neither a power law nor a thermal model alone are sufficient. While the sum of two thermal models recovers the peak at $\sim0.9$\,keV better than the best model, this model does not fit the 1-2\,keV slope. In contrast, hostW is better modeled with the sum of the thermal models ($kT=0.4$\,keV and $kT=1.3$\,keV; Figure \ref{jet_spec} middle), although the sum of a power law and a thermal model is also a good fit ($\Gamma=3.7$ and $kT=1.0$\,keV). However, the thermal model recovers the 0.7\,keV feature better. 

The host galaxy outside of the regions with the jets (hostN+hostS) is best modeled as the sum of a power law ($\Gamma=-0.4$) and an \apec thermal model ($kT=0.4$\,keV). As in HostE, fits without a power law component recover the peak better but fail to match the 1$-$2\,keV slope. The sum of two thermal models is not a significant improvement over a single thermal model, so we only include the single temperature fit in Figure \ref{jet_spec} and Table \ref{fits}. However, the likelihood of these models is low ($p=0.050$).

Photoionized emission is seen from the extended narrow-line regions (NLR) of Seyfert galaxies such as NGC 1068 and NGC 4151 \citep{ogle03, ogle00}. The nucleus of 3C\,293 is a relatively luminous for a low ionization nuclear emission-line region (LINER), so we investigate the possibility that photoionization may contribute to its extended X-ray emission. It is difficult to distinguish between thermal and photoionized plasma at the ACIS-S spectral resolution, even though they emit lines with different relative strengths.  A photoionized spectrum can be confirmed by measuring the plasma temperature via narrow recombination continuum widths with a grating spectrograph \citep{kinkhabwala02}. Lacking such data, we can test to see if the observed extended emission meets energetic and ionization requirements. The total luminosity of the host(E+W+N+S) extended emission is $2.5\times 10^{41}$\,erg\,s$^{-1}$, 2\% of the unabsorbed power-law nuclear 0.4-8 keV luminosity of $8.7\times10^{42}$ erg s$^{-1}$. Therefore, we cannot rule out significant photoionized emission by energetics alone. However, photoionization at a projected distance of 13\,kpc from the nucleus of 3C\,293 would require a very low density, insufficient to produce the observed extended X-ray flux. O {\sc vii} emission with a characteristic ionization parameter of of $\xi=$L$_{X}/n_{e} r^2 =10$ would require an electron density of $\sim6\times 10^{-4}$\,cm$^{-3}$, giving a hydrogen column density of only $N_\mathrm{H}\sim 2\times10^{19}$\,cm$^{-2}$ for a filling factor of unity, which is roughly a factor of 1000 too small to produce the observed extended soft X-ray emission. In contrast, the X-ray NLR  of NGC\,1068 has a mean column density of $N_\mathrm{H}\sim 5\times10^{22}$\,cm$^{-2}$ and mean $n_e=3.0$\,cm$^{-3}$ \citep{ogle03}.  
Moreover, from H{\sc{i}} absorption studies, \citet{beswick04} revealed column densities of $N_{H}\sim10^{21}$\,cm$^{-2}$ (assuming T$_{\rm spin}=100$\,K) for the gas in the central disk that is in front of the radio continuum. 
We can therefore rule out any significant contribution from photoionization to the extended X-ray emission in 3C 293.

\subsection{Large-Scale Jet Features}
\begin{figure}
\centerline{\includegraphics[width=\linewidth]{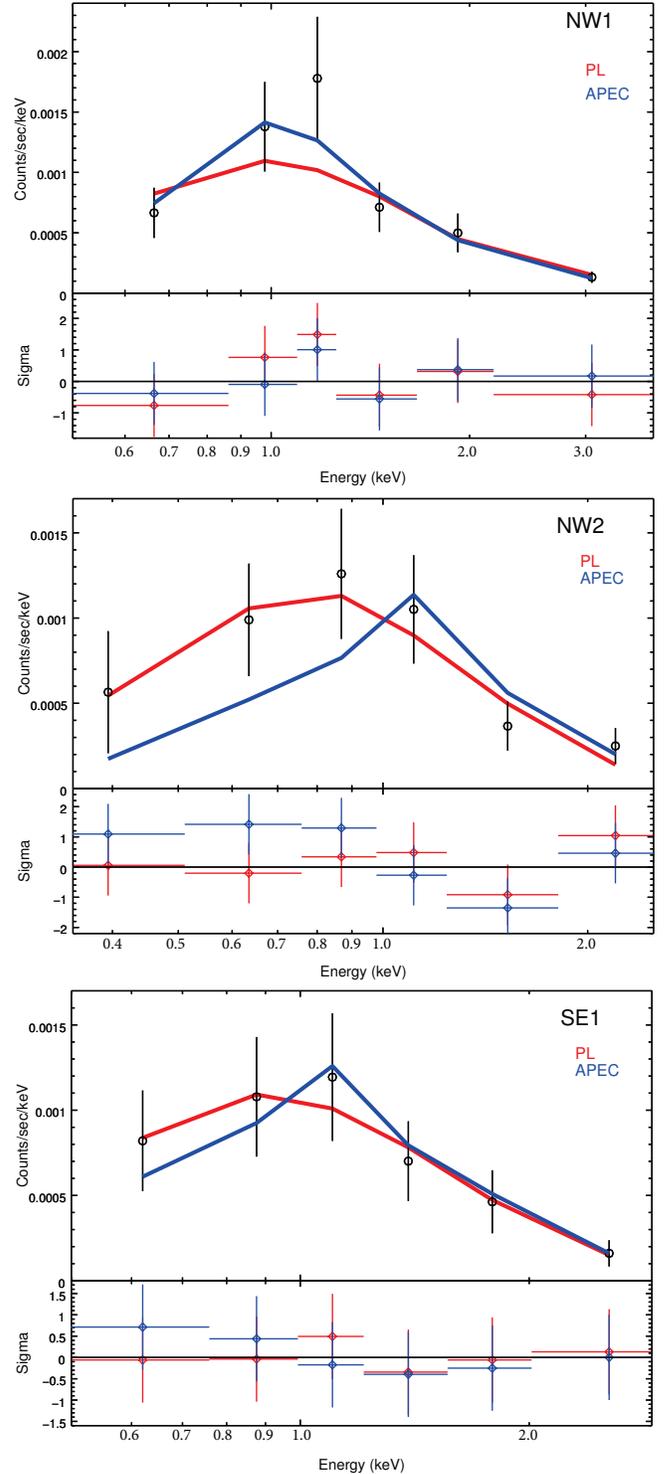}}
\caption{Spectrum of NW1 (top), NW2 (middle), and SE1 (bottom) and best-fitting model (power law (red) or \apec thermal model (blue)), with the residuals in the lower panel. }
\label{lobe_spec}
\end{figure}

The spectra of the three features along the large jets (NW1, NW2, and SE1) are shown in Figure \ref{lobe_spec}. Similar to the small-scale jets, we only fit the spectra from 0.4$-$3\,keV, which is the range over which there are significant counts. We fit each spectrum with (1) a power law alone and (2) a thermal \apec model alone, but with only six data points, there are insufficient degrees of freedom to fit more complex models (e.g. a combination of a power law and a thermal model). We impose a foreground absorber due to the Milky Way ISM and fixed the \apec model metallicity to solar. 

NW2 is best fit with a power law with $\Gamma=2.5$. A thermal model ($kT=3.3$\,keV) is significantly worse ($\chi^2$/dof=6.99/4), but it cannot be completely ruled out (probability of 0.14). In contrast, NW1 and SE1 are well fit by either a power law ($\Gamma=1.6$; $\Gamma=1.9$) or a thermal model ($kT=3.6$\,keV; $kT=4.3$\,keV).

\section{DISCUSSION}

\setcounter{table}{3}
\begin{deluxetable*}{lccclcccccc}
\tabletypesize{\scriptsize}
\tablecaption{Observed and Derived Parameters\label{parameters}}
\tablewidth{\linewidth}
\tablehead{
\colhead{Region} 	& 	\multicolumn{2}{c}{L(0.4$-$8.0\,keV)\tablenotemark{a}}	& \colhead{L(1.4\,GHz)}	& \multicolumn{4}{c}{Thermal X-ray Gas}	 \\
\cline{2-3} \cline{5-8}
\colhead{}			&	\colhead{\apec}&	\colhead{Power Law}& 	\colhead{}	& \colhead{$n_{e}$\tablenotemark{b}} & \colhead{Energy\tablenotemark{c}} & \colhead{Pressure\tablenotemark{d}} & \colhead{$\tau_{\rm cool}$\tablenotemark{e}}  \\
\colhead{}			&	\multicolumn{2}{c}{(10$^{40}$\,erg/s)}	& \colhead{(10$^{40}$\,erg/s)}	& \colhead{($f^{-1/2}$cm$^{-3}$)} & \colhead{($f^{1/2}10^{56}$\,erg)} 
	& \colhead{($f^{-1/2}10^{-11}$\,erg/cm$^{3}$)} & \colhead{($f^{1/2}10^8$\,yr)}
}
\startdata
Nucleus	&	2.3	&	39.1; 874	&	24.2~				& 	0.019~\,	&	2.2	&	4.6~\, &	4.1		\\
E. Jet	&	1.7	&	5.2		&	13.9\tablenotemark{f}	&	0.0028	&	6.6	&	0.49	&	12.~~~ 	\\
W. Jet (0.4\,keV) 	&	2.8	&	&	20.0\tablenotemark{f}	&	0.0036	&	5.9	&	0.37	&	6.7		\\
W. Jet (1.3\,keV)	&	2.6	&	&						&	0.0039	&	20.~~~ &	1.3~\,&	25.~~~	\\
Host: N+S &	1.3	&	12.2~	&						&	0.0013	&	7.4	&	0.13	&	18.~~~ 	\\
NW Jet	&		&	11.4\tablenotemark{g}	&	5.2		&			&		&		&			\\
SE Jet	&		&	5.6\tablenotemark{h}		&	1.0		&			&		&		&			
\enddata
\tablenotetext{a}{From the spectral fits in Table \ref{fits}.}
\tablenotetext{b}{Derived from \apec normalization, assuming  $n_{H}=0.8n_{e}$, uniform density, and spherical (nucleus) or fractional spherical (E/W jets) geometry. $f$ is the filling factor.}
\tablenotetext{c}{$E=3/2 n_{e} V kT$, using the \apec temperatures. $f$ is the filling factor.}
\tablenotetext{d}{$P=3/2 n_{e} kT$, using the \apec temperatures. $f$ is the filling factor.}
\tablenotetext{e}{$\tau=E/L_{\apec}$, using the \apec temperatures. $f$ is the filling factor.}
\tablenotetext{f}{From flux from \citet{beswick02}.}
\tablenotetext{g}{Sum of the NW1 and NW2 components.}
\tablenotetext{h}{From the SE1 component.}
\end{deluxetable*}

\subsection{Jet Power}

We measured the kinetic power of the jets from their radio emission in two ways. \citet{punsly05} and \citet{shabala13} each provide a means of calculating the kinetic power of the jets based on the 151\,MHz luminosity and, in the case of the \citet{shabala13} the extent of the lobes. We used the 16.08\,Jy measurement of \citet{waldram96} and a 100\,kpc extent of the lobes. These formulae give us a kinetic power of $2-4\times10^{43}$\,erg\,s$^{-1}$, which is 2$-$3 orders of magnitude larger than the radiative power at 1.4\,GHz or in the X-ray bands, and more than sufficient to heat the X-ray emitting gas.\footnote{Our value of the jet power for 3C\,293 is different from that published in \citet{guillard12} which was calculated using a version of equation 2 of \citet{punsly05} containing a typographical error. Equation 2 of \citet{punsly05} should read: $Z=3.31-3.65\times([(1+z)^4 -0.203(1+z)^3+0.749(1+z)^2+0.444(1+z)+0.205]^{-0.125})$, as given in arXiv:astro-ph/0503267 (Punsly, priv. comm.). The use of the incorrect formula results in a factor of $\sim$50 difference in the derived jet power for 3C\,293.} If this gas exists in thermal equilibrium, $\sim0.4$\% of the kinetic power needs to be dissipated to compensate for radiative losses of $1.1\times10^{41}$\,erg\,s$^{-1}$, assuming these losses are dominated by the soft thermal X-ray emission. If both the power law and thermal components of the X-ray emission observed within the host galaxy are due to dissipated kinetic power, then the amount needed rises to $\sim1-2$\%.

\subsection{ISM Emission}

\subsubsection{Hot ISM}

We use the \emph{Chandra} spectra to characterize the X-ray emitting gas in the host galaxy, which may be heated by the radio jets. From the thermal components of the best fit of the nuclear and host spectra, we calculate the electron density, energy, and cooling timescale of the thermal plasma. The electron density can be calculated directly from the normalization of the \apec models (see Table \ref{fits}). We assume constant density in taking the volume integral and $n_{H}=0.8 n_{e}$, which corresponds to a fully ionized plasma with 10\% He. We set the nuclear volume to be a sphere with a radius the size of the extraction aperture and the host volumes to be segments of the 20$''$ sphere corresponding to their extraction region (e.g. V$_{\rm East}=(4/3)\pi(20'')^3\times (55/360)$). The resulting electron densities, given in Table \ref{parameters}, range between $1.3\times10^{-3}\, {f^{-1/2}\,cm^{-3}}$ and  $1.9\times 10^{-2}\, {f^{-1/2}\, cm^{-3}}$, which are reasonable for hot ISM.\footnote{Here and below, $f$ is the filling factor.} Based on the thermal emission, we also measure the mass of X-ray emitting gas\footnote{M$_{X}=f\,V\,m_{p}\,n_{H}$} (Table \ref{ism}). We find masses of (5, 2, and 8)$\times10^{8}{f^{1/2}}$\,M$_{\odot}$ of hot gas in the North+South, East, and West host regions,  cooling at a rate of 1.1\,M$_{\odot}$\,yr$^{-1}$. 

\begin{figure*}
\centerline{\includegraphics[width=\linewidth]{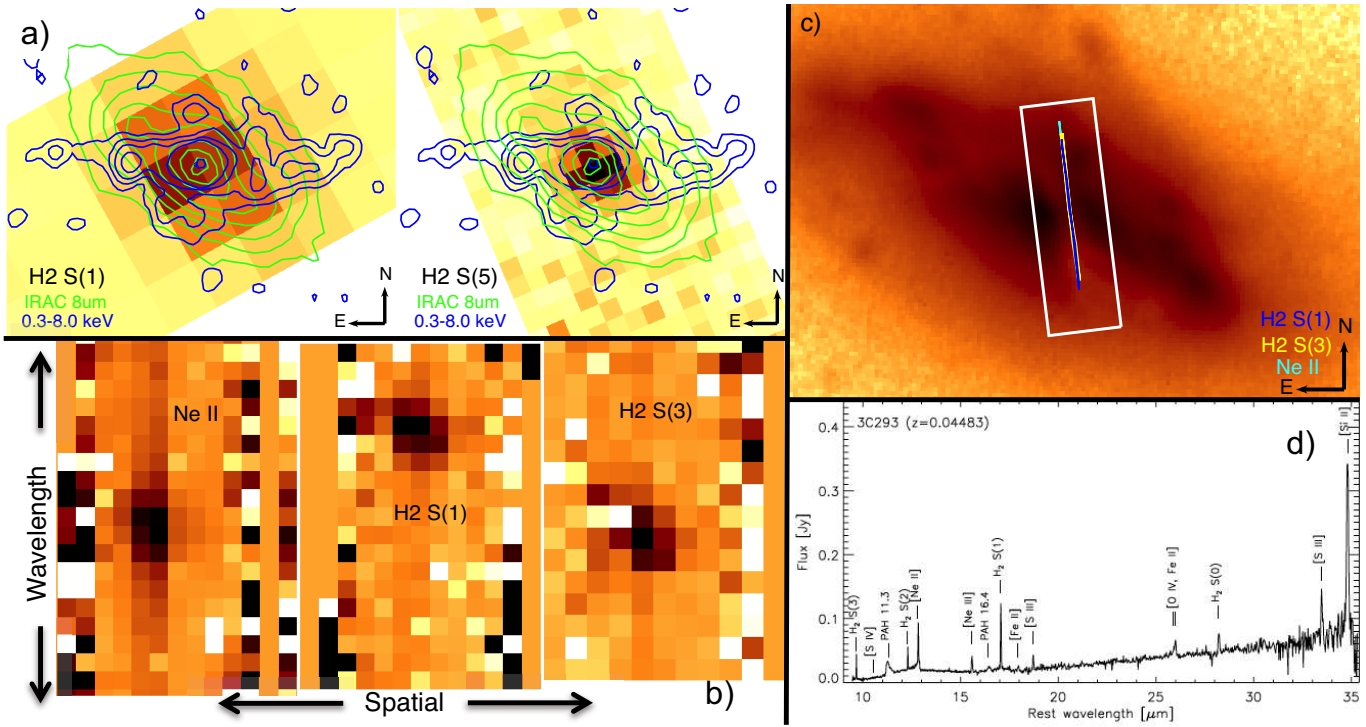}}
\caption{IRS maps of 3C\,293 maps (a) of the \mh S(1) and \mh S(5) emission overlaid with contours of 8\um (green) and X-ray (blue) emission showing that the \mh emission is concentrated near the nucleus, but extends over much of the galactic disk. Cut-outs of the staring observation (b) at the \mh S(1), \mh S(3), and [Ne {\sc{ii}}] lines similarly show that the shocked gas is found over much of the galactic disk (c; B-band image also shown in Fig. \ref{spitzer}). The integrated spectrum, previously published in \citet{guillard12}, is shown in panel (d).}
\label{irsmap}
\end{figure*}

\subsubsection{Warm \mh Emission}

\citet{ogle10} measured a total luminosity of the \mh rotational lines $0-0~S(0)-S(3)$ of $5.75\times10^{41}$\,erg\,s$^{-1}$. A small percentage ($\sim$1\%) of kinetic energy dissipated from the jets is more than sufficient to account for the observed warm \mh emission. The $3.7\times10^{9}$\,M$_{\odot}$ of warm molecular hydrogen at 100\,K observed by \citet{ogle10} contains 5$\times10^{52}$\,erg of thermal energy, about 4 orders of magnitude smaller than the $\sim10^{56-57}$\,erg contained in the hot ISM, calculated based on its temperature and electron density.

Figure \ref{irsmap} shows the extent of the warm ISM with respect to the thermal X-ray emission. There is a concentration at the center of the galaxy, seen in particular in the H2 S(5) map, but warm \mh is detected over much of the galactic disk, as shown by the \mh S(1) map (Fig. \ref{irsmap}a). The staring observation also allows us to determine that the [Ne {\sc ii}], which can also be heated by shocks, is likewise found over most of the length of the SH slit, which covers much of the breadth of the galactic disk.

\begin{deluxetable}{llccccl}[b]
\tabletypesize{\scriptsize}
\tablecaption{Luminous Galaxy Contents\label{ism}}
\tablewidth{0 pt}
\tablehead{
\colhead{Component} & \colhead{Mass (M$_{\odot}$)} 	&	\colhead{Temperature}		}
\startdata
Hot ISM\tablenotemark{a}		& 	$1.6\times10^{9}$	&	0.4-1.3\,keV	\\	
Warm \mh\tablenotemark{b}	&	$3.7\times10^{9}$	&	100-1040\,K	\\	
Ionized Gas\tablenotemark{c} 	& 	$7~~\,\times10^{5}$	& 				\\
Cold \mh\tablenotemark{d}	&	$2.2\times10^{10}$	&				\\
Dust\tablenotemark{e}		&	$6.7\times10^{7}$	&	20\,K and 59\,K \\ 
Stellar\tablenotemark{e}		&	$1.7\times10^{11}$	&			
\enddata
\tablenotetext{a}{Hot M$_{H}$ calculated based on the thermal X-ray components of host region spectra.}
\tablenotetext{b}{From \citet{ogle10}.}
\tablenotetext{c}{Mass measured from the narrow and broad H$\beta$ lines in the E0 region \citep{emonts05}.}
\tablenotetext{d}{From \citet{evans99} and \citet{labiano13}.}
\tablenotetext{e}{Based on the SED fit.}
\end{deluxetable}

\subsubsection{Dust and CO Emission}

\begin{figure}
\centerline{\includegraphics[width=\linewidth]{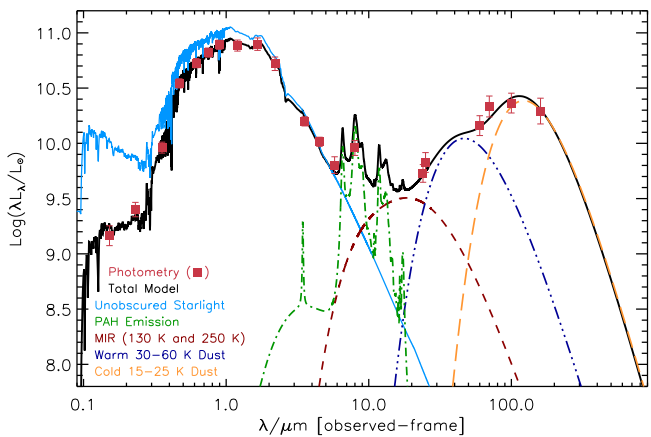}}
\caption{Integrated SED of the host of 3C\,293, using GALEX, SDSS, 2MASS, IRAC, MIPS, and IRAS photometry, fit with \magphys. The fitted model requires a stellar mass (blue) of $1.7\times10^{11}$\,M$_{\odot}$, and a dust mass of $6.7\times10^{7}$\,M$_{\odot}$ primarily at 20\,K (orange) and 59\,K (purple). The SED was extracted in the 30$''$ circular aperture shown in Figure \ref{spitzer}. }
\label{sed}
\end{figure}

Figure \ref{sed} shows the spectral energy distribution (SED) of the host galaxy measured in a 30$''$ aperture from the UV to FIR. We used the SED modeling code \magphys \citep{dacunha08} to fit it and derive properties of the host galaxy. \magphys fits SEDs with a stellar spectra library derived from the \citet{bruzual03} stellar population synthesis code and an infrared dust spectrum, comprised of a polycyclic aromatic hydrocarbon (PAH) template, a MIR continuum at fixed temperatures (130\,K and 250\,K) and two thermal dust components. The SED fit estimates the dust mass of 3C\,293 at $7\pm3\times10^{7}$\,M$_{\odot}$. However, \citet{lanz13} and \citet{aniano12} both noted that SED fits done in the absence of photometry at $\lambda \ge 170\mu$m, as is the case here, tend to underestimate the cold dust temperature and hence overestimate the dust mass by as much as 60\%.

3C\,293 was the first FRII galaxy in which CO was detected \citep{evans99}. More recent observations by \citet{labiano13} determined that CO is distributed in a highly structured, elongated disk extending out to 12$''$ from the nucleus (Fig. \ref{spitzer}). These observations also suggest the presence of a CO absorber in front of the AGN, interpreted by \citet{evans99} and \citet{labiano13} as circumnuclear clouds on noncircular trajectories. 3C\,293 contains a total cold \mh mass of $M(\mhb)=2.2\times10^{10}$\,M$_{\odot}$ \citep{evans99,labiano13}\footnote{Mass was derived assuming a \mh-mass to CO-luminosity ratio of 4.6\,M$_{\odot}$/K\,km\,s$^{-1}$\,pc$^{2}$ \citep{solomon87}.}. Therefore, the gas-to-dust ratio in 3C\,293 is $330$. If the dust mass is indeed overestimated due to the limited FIR observations, then this ratio may be even higher. A similar effect is seen in the MOHEG NGC\,4258, which has an even greater CO-inferred gas-to-dust ratio of 1000 \citep{ogle14}, as well as in the Taffy bridge, which shows an enhancement level similar to 3C\,293 \citep{zhu07}. However, shocks may enhance the CO emission in 3C\,293 (as suggested by the highly excited CO emission found by \citet{papdopoulos08}), resulting in an over-estimation of the \mh mass, if a typical $X_{CO}$ is assumed, as well as the related gas-to-dust ratio. Table \ref{ism} also shows that 3C\,293 has a warm-to-cold \mh ratio of $\sim0.17$.

\subsection{Jet Feedback on the ISM}
\subsubsection{X-ray Emission in Radio Galaxies}

\begin{deluxetable*}{lcccclccl}
\tabletypesize{\scriptsize}
\tablecaption{Comparison Among Galaxies\label{glx}}
\tablewidth{\linewidth}
\tablehead{
\colhead{Galaxy} & \multicolumn{5}{c}{X-ray Properties} 	&\multicolumn{3}{c}{IR Properties}  	\\
\cline{2-6} \cline{7-9}
\colhead{} & \colhead{log(L$_{\rm X,\,AGN}$}&\colhead{log(L$_{\rm Thermal}$}&\colhead{E$_{\rm Thermal}$}&\colhead{T}&\colhead{Ref.\tablenotemark{a}} & \colhead{log(L$_{\rm24\mu m,\,AGN}$}&\colhead{log(L$_{\mhb}$}	& \colhead{Ref.\tablenotemark{b}}\\
\colhead{} & \colhead{(erg\,s$^{-1}$))}	& \colhead{(erg\,s$^{-1}$))}	&	\colhead{(erg)}	&	\colhead{(keV)}	 & & \colhead{(erg\,s$^{-1}$))} &\colhead{(erg\,s$^{-1}$))} 	& 
}
\startdata
3C\,293		& 42.9	& 41.0	&	7$\times10^{56}$		& 0.7, 1.0	&1	& 43.3	& 41.8		& 12	\\
NGC\,4258	& 40.9	& 39.9	&	5$\times10^{55}$		& 0.5		&2, 3& 42.7	&  39.9		& 1, 13 	\\	
Cen\,A 	 	& 41.7	& 40~~ 	&	2$\times10^{56}$		& 0.3		&4, 5& 42.7	&  39.9		& 12	\\	
3C\,305		& 41.2 	& 40.3	&	6$\times10^{56}$		& 0.8		&6, 7&  43.4	& $<$41.4~~\,	&  7   	\\	
3C\,321		& 43.6	& 40.7	&	...					& 0.3		&8	&  44.9	&  $<$42.1~~\,	& 12, 14	 \\	
PKS\,1138-26   & 45.6	& 45.1	&	2$\times10^{61}$		& 5~~\,		&1,9	&...		& 44.9	& 1, 15	 \\
NGC\,3801	& 41.5	& 39.8	&	5$\times10^{55}$		& 0.2-1.0  &10	&  ...	& ...				& ...		 \\
4C\,29.30		& 43.7	& 42.0	&	$10^{56-57}$			& 0.5		&11	&  ... 		& ...			&	...
\enddata
\tablenotetext{a}{References for the X-ray properties: (1) This paper, (2) \citet{yang07}, (3) \citet{young04}, (4)  \citet{kraft00}, (5) \citet{kraft03}, (6) \citet{hardcastle12}, (7) \citet{guillard12}, (8) \citet{evans08}, (9) \citet{carilli02}, (10)  \citet{croston07}, and (11)  \citet{siemiginowska12}}
\tablenotetext{a}{References for the IR properties: (1) This paper, (7) \citet{guillard12}, (12) \citet{ogle10}, (13) \citet{ogle14}, (14) \citet{Dicken10}, and (15) \citet{ogle12}. }
\end{deluxetable*}

The presence of thermal X-ray gas associated with radio jets has been identified in a number of radio galaxies (e.g., NGC\,4258: Yang et al. 2007; Centaurus A: Kraft et al. 2009; 3C\,321: Evans et al. 2008). \nocite{yang07, kraft09, evans08} In several such galaxies, X-ray emission is seen to trace the radio jets but with a wider, resolved region of emission. \citet{wilson01} describe the X-ray emission, which \citet{yang07} determined to be thermally emitting gas at $\sim0.5$\,keV, as ``enveloping'' the radio jets that rise to the anomalous arms of NGC\,4258. Similarly, \citet{massaro09} argue that the diffuse X-ray emission in 3C\,305 is observed to surround the radio emission rather than exist solely coincident with it, attributing this thermal emission to gas collisionaly heated by shocks driven by the jets.  

3C\,305, particularly, has similar morphological and spectral X-ray properties to 3C\,293 (Table \ref{glx}). Both show extended, thermal X-ray emission along the direction of the radio jets out to $\sim$10\,kpc from the nucleus. 4C\,29.30 \citep{siemiginowska12} and NGC\,3801  \citep{croston07, emonts12} likewise show large quantities of thermal X-ray emission associated with the radio jet propagating through their host galaxies. At higher redshift, the Spiderweb galaxy \citep[PKS1138-26, z=2.16; ][]{carilli02, ogle12} contains a thermal X-ray cocoon around its radio jets and is an extremely luminous MOHEG. The situation is further complicated in 3C\,293 where radio observations \citep[e.g.,][]{beswick02, joshi11} have captured either the large  $\sim100$\,kpc jets or the small $\sim5$\,kpc jets. However, the flux contained within the small scale radio jets appears to account for the unresolved flux in the core when the large jets are resolved. Therefore, it appears that the X-ray jets do continue $\sim10$\,kpc beyond the inner radio jets, in a  region devoid of bright radio emission. This mismatch between the X-ray and radio jets may be a result of a longer cooling timescale for the X-ray emitting cocoon compared to the synchrotron-emitting jets or the faster motion of the radio features in the jet compared to the cocoon.

Observations of more nearby systems with better spatial resolution including NGC\,4258 and Centaurus A have produced the picture \citep[e.g.,][]{wilson01, kraft07} that radio jets through their interaction with the ISM shock gas to X-ray emitting temperatures.  For example, the shell of hot (3.5\,keV) gas detected around the SW lobe of Centaurus A \citep{kraft03, kraft07} was interpreted as the result of a supersonic expansion of the radio lobe into the surrounding ISM. However, the jet of Centaurus A is much more perpendicular to its host galaxy. NGC\,4258, in contrast, has a similar geometry to 3C\,293 in that the radio jets initially traverse the denser galactic disk. In the process, they shock the ISM gas. Once the jets leave the main galactic disk, they continue to drive mass motions and shocks through the lower density ISM, creating an expanding cocoon of shocked gas \citep{sutherland07, wagner12}. As a result, while the solid angle of the jet remains small, its area of impact can eventually cover the whole host galaxy. \citet{wilson01} suggest that the diffuse emission they measure around the anomalous arms of NGC\,4258 is primarily the result of the mass motions directed towards the galactic disk hitting and shocking the disk ISM to X-ray temperatures. This hot gas is also blown out of the disk as a result, creating the emission along the anomalous arms that is particularly bright on the edge closest to the disk. 

Applying this picture to 3C\,293 implies that the emission observed east and west of the nucleus is likely the result of ISM shocked to X-ray temperatures. The large radio jets could drive mass motions into the galactic disk, shocking gas to further contribute to the observed emission. Only a tiny fraction of the kinetic power of the radio jets ($<1$\%) would be necessary to account for the X-ray emission. The morphology of the X-ray emission within the host of 3C\,293  further supports this picture, as it hints at similarities to the anomalous arms of NGC\,4258. The western jet in particular has a suggestion of a curve towards the larger NW jet, reminiscent of the departure of the anomalous arms of NGC\,4258 from the disk of their host galaxy.

\subsubsection{Relation of \mh and Diffuse X-ray Emission}

\begin{figure}
\centerline{\includegraphics[width=\linewidth]{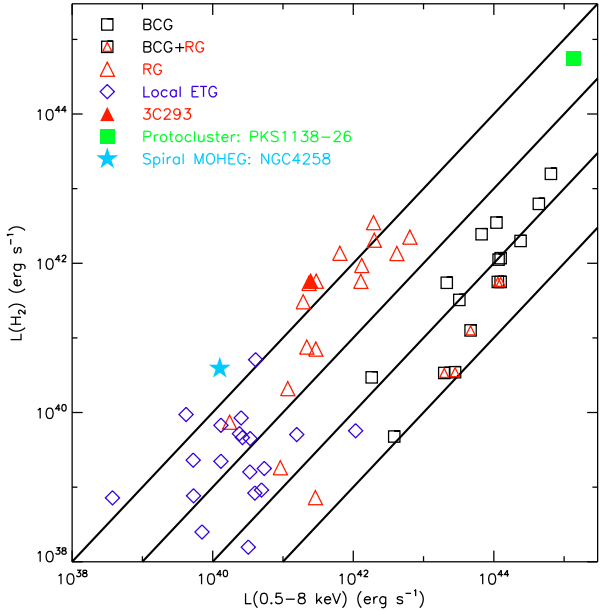}}
\caption{\mh luminosity compared to X-ray (0.5-8\,keV) luminosity of the host MOHEG galaxies excluding the AGN. The black lines show where L$_{\mhb}$/L$_{X}$=[1, 0.1, 0.01, 0.0001]. Radio galaxies (red triangles) generally have similar  L$_{\mhb}$ and L$_{X}$, with higher L$_{X}$ than the typical early type galaxy (ETG; blue diamonds). The filled symbol is 3C\,293. Brightest cluster galaxies (BCGs; black squares) have X-ray luminosity associated with the cluster within the host galaxy, so their L$_{X}$ is approximately two orders of magnitude larger than their  L$_{\mhb}$.}
\label{h2lx}
\end{figure}

In this picture, the shocks driven by the jet power both the thermal X-ray emission and the warm \mh emission. Figure \ref{h2lx} shows the relation between L$_{\mhb}$ and the X-ray luminosity of MOHEGs hosted in radio galaxies, brightest cluster galaxies (BCGs) and local early type galaxies (ETGs). The appendix describes how the X-ray luminosity was extracted in detail. In brief, we extracted spectra in the aperture given by the stellar extent, excluding the central hard point source. We find that radio galaxies, including 3C\,293, which are not BCGs, have similar L$_{\mhb}$ and L$_{X}$. In contrast, the X-ray luminosity of BCGs is about two orders of magnitude larger than  L$_{\mhb}$. We suggest that in the non-BCG radio galaxies both the X-ray and \mh emission is powered by shocks driven by the jet, whereas in BCGs, the X-ray emission is dominated by cooling in the gravitational potential that has created a well-developed hot atmosphere over periods much longer than the current jet activity. ETGs have a wide range of X-ray luminosities relative to their L$_{\mhb}$.\footnote{One caveat is that the local ETGs are closer than the radio galaxies and BCGs and hence the IRS slit only covered a portion of the galaxies, thereby possibly underestimating the L$_{\mhb}$ of these galaxies.} 

We suggest the following evolutionary interpretation for the locations of different types of galaxies in the plot. When the radio jet is activated, perhaps by a merger, the interaction of the jet with the ISM drives the galaxy up and to the right in Figure \ref{h2lx}, as it drives increases in both \mh and X-ray luminosity. Without energy injection, the \mh cools more quickly than the hot ISM, moving galaxies downward in the plot. However, as the \mh emission is dependent on the dissipation of kinetic energy, its cooling occurs on the turbulent dissipation timescale, which can be longer than the time between two active jet phases, as has recently been shown for 3C\,326 (Guillard et al. 2014, in press.). The net effect of each cycle of radio jet activity is to drive a galaxy to the right (higher X-ray luminosity). 

From the ratio of the \mh and diffuse, thermal X-ray luminosities, we can extract the ratio of their masses and filling factors ($f$), assuming both gases fill the same volume V. 
 The \mh luminosity is given by:
\begin{eqnarray}
L_{\mhb} &=& \frac{M_{\mhb}}{2 m_{p}}  X_{\mhb} = f_{\mhb} V  n_{\mhb} X_{\mhb}{\rm, where~~ } \\
X_{\mhb} &=& 3kT_{\mhb} / \tau_{\mhb}~.	 
\end{eqnarray}
The bulk of 3C\,293 warm \mh mass ($M_{\mhb}$) is at a temperature ($T_{\mhb}$) 100\,K \citep{ogle10}, where the mean luminosity per molecule $X_{\mhb}=4.7\times10^{-26}$\,erg/s, and hence the cooling timescale is $\tau_{\mhb}=3\times10^{4}$\,yr, assuming no constant energy injection into the \mh gas. 
 The thermal X-ray luminosity is given by: 
\begin{eqnarray}
L_{X} &=& \frac{M_{X}}{m_{p}} n_{e} X_{X} =  f_{X} V  n_{e}^{2} X_{X}{\rm , where} \\
X_{X} &=& 3/2~kT_{X}/( n_{e}  \tau_{X})~.
\end{eqnarray}
For the typical temperature $T_{X}$ (0.8\,keV) and $n_{e}$ ($3\times10^{-3}$\,cm$^{-3}$) of the hot plasma (Table \ref{parameters}), log($X_{X}/({\rm erg~s^{-1}~cm^{3}}))=-22.5$ \citep{allen00} and $\tau_{X}=6\times10^{9}$\,yr. 
  The ratio of the \mh and X-ray luminosities then yields the ratio of filling factors and masses of each gas phase: 
\begin{eqnarray}
\frac{L_{\mhb}}{L_{X}} = \frac{n_{\mhb}}{n_{e}^2}\frac{f_{\mhb}}{f_{X}}\frac{X_{\mhb}}{X_{X}} = \frac{M_{\mhb}}{M_{X}} \frac{X_{\mhb}}{2 n_{e} X_{X}}~.
\end{eqnarray}
We estimate $n_{\mhb}$=$1\times10^{3}$\,cm$^{-3}$ (Guillard et al. in prep.), so 3C\,293's luminosity ratio of 2.33 yields $f_{\mhb}/f_{X}=1\times10^{-5}$ and  $M_{\mhb}/M_{X}=9$,  within a factor of a few of the ratio of the masses given in Table \ref{ism}. Assuming similar densities, radio galaxies, which typically have $L_{\mhb}/L_{X}\sim1$, have $f_{\mhb}/f_{X}\sim10^{-5}$ and $M_{\mhb}/M_{X}\sim4$. The similarities in the luminosities and masses of warm \mh and thermal X-ray gas are consistent with the picture that in these systems both derive from a multiphase, shocked ISM. In contrast, BCGs whose $L_{\mhb}/L_{X}\sim0.01$ have $f_{\mhb}/f_{X}\sim10^{-7}$ and $M_{\mhb}/M_{X}\sim0.04$, because their massive, virialized, hot atmospheres overwhelm the X-ray emission due to the current jet-ISM interaction. As an intermediary case, the Spiderweb BCG (PKS1138-26), a powerful radio source that resides in an unvirialized protocluster, has $L_{\mhb}/L_{X}\sim0.4$. 

The X-ray cooling timescale ($\tau_X$) is $\sim10^{5}$ times longer than $\tau_{\mhb}$, so shock-heated X-ray plasma has more time to spread from the shock sites and fill more of the emitting volume before cooling than the \mh emission. Over time, jet-ISM interactions may contribute to the establishment of a hot atmosphere in massive ellipticals including BCGs. If the jet operates for 10$^{7}$\,yr to generate the observed X-ray luminosity in 3C\,293, then it needs to operate either continuously for $10^{9}$\,yr or with a duty cycle of 10\% for $10^{10}$\,yr to generate the X-ray luminosity observed in BCGs. If the energy used by the jet to drive outflows also dissipates in the host, then this process may be even more efficient at generating the hot atmosphere. Alternatively, galaxy interactions, including minor mergers, may also contribute to the buildup of the X-ray atmospheres.

\subsubsection{Gas Outflows}

Jet feedback on ISM can also drive outflows. 3C\,293 is known to house both ionized and neutral gas outflows \citep{morganti03, emonts05, mahony13}. In particular, radio observations have detected a broad ($\sim1200$\,km\,s$^{-1}$) neutral HI outflow from the small-scale western radio jet \citep{mahony13, morganti03} and optical spectroscopy detected a fast-moving ionized outflow primarily associated with the small-scale eastern jet \citep{emonts05}. These authors have argued that the jets of 3C\,293 are capable of driving these outflows. This is further supported by the application of the work of \citet{wagner12} to 3C\,293. \citet{wagner12} used hydrodynamical simulations to determine that radio jets are capable of accelerating ISM clouds to the velocities found in the outflows of 3C\,293 by sweeping up material ablated by ram pressure of high-velocity flows in the channels through a porous ISM, provided that the ratio of jet power to Eddington luminosity $\eta=P_{\rm jet}/L_{\rm Edd.}\gtrsim10^{-4}$. In 3C\,293, our measurement of the jet power yields $\eta=0.002$, more than sufficient to accelerate clouds to outflow velocities.

Figure \ref{chandra_vla_nuc} shows the region in which these outflows are being detected. Although the \chandra resolution is significantly coarser than the resolution in the radio, we detect significant soft X-ray counts coincident with Region E0 of \citet[][our region NE0]{emonts05}. E0 contains a blue shifted, broad-component emission-line feature seen in [S {\sc{ii}}] and  [O \textsc{ii}] (with FWHM $\sim730$\,km\,s$^{-1}$ and $\Delta$v$\sim-540$\,km\,s$^{-1}$ with respect to the narrow-component emission-line disk), which was interpreted as the base of a fast, ionized outflow. However, the estimate of \citet{emonts05} of the spatial extent of this outflow of $2\times1.5$\,kpc$^2$ is larger than the extent of the soft X-ray emission, suggesting that we are detecting the X-rays only at the launch site of the outflow.  In contrast, the neutral outflow, which launches from the knot in the inner western jet \citep{beswick04}, is closer to the AGN and hence, we cannot determine with the resolution of this observation whether it too has X-ray emission associated with its launch site. The absence of such emission might suggest that the same process responsible for the heating the X-ray emitting gas is ionizing the eastern outflowing gas. 

The kinetic power of the jet ($\sim(2-4)\times10^{43}$\,erg\,s$^{-1}$; Sect. 5.1) is likely sufficient to power these outflows, even if this path of energy dissipation did not simultaneously help to power the X-ray or \mh emission. However, a much larger fraction of the kinetic power is needed to drive these outflows than to power the emission. The ionized outflow is estimated to dissipate $\sim10^{56}$\,erg over its 10$^{6}$\,yr lifetime, requiring an average $\sim10^{42}$\,erg\,s$^{-1}$ to drive it \citep{emonts05}, while the energy loss rate of the more massive neutral outflow estimated by \citet{mahony13}  is between $1.4\times10^{42}$\,erg\,s$^{-1}$ and $1.0\times10^{43}$\,erg\,s$^{-1}$ based on a mass-flow rate of 8-50\,M$_{\odot}$\,yr$^{-1}$. The upper end of the mass flow and energetics estimates of the neutral flow are derived assuming that the entire broad H{\sc{i}} component is due to an outflow at 600\,km\,s$^{-1}$. \citet{mahony13} argue that a plausible alternative scenario has part of this absorption profile being due to turbulent gas in a rotating disk, resulting in a conservative lower limit of 8\,M$_{\odot}$\,yr$^{-1}$ flowing out at 100\,km\,s$^{-1}$ with an energy loss rate of $1.4\times10^{42}$\,erg\,s$^{-1}$. This scenario is also similar to that proposed by \citet{beswick04} for the intermediate H{\sc{i}} seen against the inner western lobe. The energetics of the jet compared to the outflow also argue for a neutral mass outflow on the lower end of the range given by \citet{mahony13}. The diffuse X-ray luminosity provides a gauge of the amount of energy dissipated into the host galaxy. At \.{M}$=8$\,M$_{\odot}$\,yr$^{-1}$, driving the neutral outflow only requires $\sim5$\% of the jet's kinetic power ($\sim10^{42}$\,erg\,s$^{-1}$), only a factor of $\sim4$ larger than the diffuse X-ray luminosity. In contrast, if the neutral outflow has an \.{M}$=50$\,M$_{\odot}$\,yr$^{-1}$, then the energy necessary for the jet to drive such an outflow ($\sim10^{43}$\,erg\,s$^{-1}$) would sap much of the kinetic power of the jet.

\subsection{Extended Structures}

The large-scale radio morphology of 3C\,293 is that of an FR\,\textsc{ii} with lobes of unequal surface brightness. Its associated X-ray features are reminiscent of features seen in Cen A. In particular, the SE1 and NW2 components are similar to the shell of X-ray emission around the southern lobe of Cen A \citep{kraft03}. NW1 has some similarity with the thermal knots detected in Cen A's northern middle lobe \citep{kraft09}, but may be the X-ray hotspot associated with the radio peak (Figure \ref{chandra_aps}c).

There are several possible explanations for the X-ray emission from these structures along the large-scale radio jets. It could be direct synchrotron emission, synchrotron self-compton (SSC), or inverse-Compton scattering of cosmic microwave background (IC-CMB) photons off of the radio-synchrotron emitting relativistic electrons of the lobes. Alternatively, it could be the result of the jet colliding with intergalactic medium or material dragged out of the host galaxy in an outflow during a previous outburst. However, it does not appear likely that the jet is colliding with material from the tidal tails. Based on the extended emission seen in the B-band (Figure \ref{spitzer}), the bulk of tidal material is roughly perpendicular to the axis of the radio jets in the plane of the sky. 

The spectra of the three regions (Fig. \ref{lobe_spec}; Table \ref{fits}) are consistent with a power law model, making synchrotron, SSC, and IC-CMB emission plausible explanations. The extrapolation of the radio flux, assuming $\alpha=1$, to the predicted X-ray flux due to direct synchrotron emission, yields a ratio of observed-to-predicted X-ray flux of $2-4$ for NW1 and NW2, placing these features in the regime of hotspots where \citet[][see esp. their Fig. 2]{hardcastle04} argued that a synchrotron model is natural. The upper limits from \emph{Spitzer} and SDSS data do not provide interesting constraints on the synchrotron model. For all three features, the photon energy density is smaller by a large factor ($>$1000) than the energy density in the magnetic fields \citep{joshi11}, indicating that the majority of the energy loss happens through synchrotron emission rather than SSC. Given the low-redshift of 3C\,293, IC-CMB emission is likewise unlikely to contribute significantly to the observed emission without important beaming effects, which are not indicated by the morphology of 3C\,293.

NW1 fits well with either a power-law model or a thermal model. If its emission is due to synchrotron emission, a plausible scenario as described above, then it may be associated with the radio hotspot 6.6\,kpc away (Fig. \ref{chandra_aps}c). Assuming a speed of 0.15$c$ \citep{harwood13}, the distance between the peak of the X-ray and radio emission would require $\sim10^5$\,yr to cross. This is long compared to the synchrotron lifetime for electrons emitting in the X-ray regime ($\sim10^3$\,yr in the 11\,$\mu$G magnetic field estimated by \citet{joshi11}), but short compared to the synchrotron lifetime for electrons emitting in the radio ($\sim10^7$\,yr). The longer-lived radio-emitting particles are visible downstream, at a distance that the X-ray emitting particles could not travel in their lifetime. Offsets between radio and X-ray emission have been seen in other radio galaxies (e.g. 3C\,390.3 in Hardcastle et al. 2007 and 3C\,351 in Hardcastle et al. 2002), and it has been argued that such spatial differences cannot be accounted for with SSC or IC-CMB models without requiring implausibly large spatial variations in the low-energy electron populations and/or important beaming effects. \nocite{hardcastle02, hardcastle07} 

If the emission in NW1 is thermal instead, then it requires a fairly high temperature of 3.6\,keV to explain its emission, and it is morphologically similar to the thermal knots found in the northern middle lobe of Cen A, which were determined by \citet{kraft09} to be most likely due to a jet-cloud interaction resulting in shock-heating of cold gas. However, the Cen A knots are noticeably cooler with temperatures between 0.4\,keV and 1\,keV \citep{kraft09}.

In contrast, SE1, and to a lesser degree NW2, have morphologies more reminiscent of a shell or cap of shock-heated X-ray gas around a supersonically expanding radio lobe. SE1 is particularly morphologically reminiscent of the X-ray emission detected along the edge of the southern lobe of Cen A \citep{kraft03}. There is also some morphological similarity to the curved feature on the edge of the northern lobe of 4C\,29.30 \citep[EArm; ][]{siemiginowska12}, which, like the Cen A feature, is thermal in origin. Although it does not appear on the end of the SE lobe, Figure \ref{chandra_aps}a shows that stronger radio emission is present on either side of SE1 along the lobe, suggesting this feature may lie at the forefront of an ejection burst, perhaps associated with a bowshock. The spectrum of SE1 is equally well fit with a power-law or a thermal model. In contrast, NW2 is more likely to be synchrotron emission ($\Gamma=2.5$, p=0.68) but we cannot rule out a thermal model (p=0.14). Without a deeper observation, however, we cannot better determine the origin of the emission in these features.

\section{SUMMARY}

We presented a 70\,ks \chandra observation of the radio galaxy 3C\,293, which shows evidence for jet-shocked, X-ray emitting gas both within the galaxy and at several locations outside of the galaxy. We find the following:
\begin{enumerate}
\item In addition to hard emission from the AGN, the nuclear region contains a softer off-nuclear source associated with the launch site of the ionized outflow. 
\item Most of the X-ray emission within the host galaxy has jet-like morphology to the East and West of the nucleus, which is best fit with two thermal models (West) and the combination of a thermal model and a power-law model (East).
\item Diffuse, soft X-ray emission is detected to the North and South of the AGN within the host galaxy and requires both a thermal and power-law component to fit its spectrum.
\item We detect two curved features along the large radio jets: SE1 is reminiscent of the thermal X-ray emission around Cen A's South Lobe and the north lobe of 4C\,29.30, while NW2 is more likely to have synchrotron emission.
\item We also detect a possible X-ray counterpart to the NW jet hotspot, whose emission is plausibly explained by synchrotron emission.
\end{enumerate}

The morphologies and luminosities of 3C\,293 fit well into a picture whereby the jet affects the ISM through shocks. The kinetic power of the jet is more than sufficient to power the X-ray thermal emission and the warm \mh emission as well as driving both the ionized and neutral outflows found in this system. As the jet crosses its host galaxy, it drives shocks into the ISM, heating some of the gas up to X-ray emitting temperatures and creating a cocoon of hot gas. This hot gas then expands in a more spherical manner, thereby bringing the effects of the radio jet to a much larger portion of the host galaxy than the cross-section of the jet itself, and could transfer energy to the \mh either by driving shocks into dense clouds or through turbulent mixing. Only a small fraction of the kinetic energy of the jets ($<3$\%) needs to be dissipated to account for both the thermal X-ray and \mh emission, well within the range of the efficiency of energy transfer expected from numerical simulations.

 We compared the diffuse X-ray luminosity and warm \mh luminosities in a sample of MOHEGs housed in radio galaxies, BCGs, and local ellipticals. We find that radio galaxies, including 3C\,293, have a ratio of \mh and X-ray luminosities of $\sim1$. This is consistent with the picture that the shocks driven into  the ISM by the radio jet powers both types of emission in these systems. In contrast, BCGs, housed in deeper gravitational wells with longer lasting X-ray halos have $L_{\mhb}/L_{X}\sim0.01$. However, multiple jet-ISM interactions may serve to help build up a hot, virtualized atmosphere in these giant galaxies. 

\acknowledgements L. L. thanks Aneta Siemiginowska and Katherine Alatalo for insightful discussions. We also thank Alvaro Labiano for sharing the CO data from his paper. Support for this work was provided by the National Aeronautics and Space Administration through Chandra Award Number GO1-12122X issued by the Chandra X-ray Observatory Center, which is operated by the Smithsonian Astrophysical Observatory for and on behalf of the National Aeronautics Space Administration under contract NAS8-03060. The scientific results reported in this article are based on observations made by the Chandra X-ray Observatory and data obtained from the Chandra Data Archive, some of which was published previously in cited articles. This work also used archival data obtained from the Spitzer Science Archive, the Mikulski Archive for Space Telescopes (MAST), and the NASA/ IPAC Infrared Science Archive (IRSA). \emph{Spitzer} is operated by the Jet Propulsion Laboratory, California Institute of Technology under a contract with NASA. \emph{GALEX} is operated for NASA by the California Institute of Technology under NASA contract NAS5-98034. This research has made use of the NASA/IPAC Extragalactic Database (NED), which along with IRSA, is operated by the Jet Propulsion Laboratory, California Institute of Technology, under contract with the National Aeronautics and Space Administration. B. E. acknowledges funding through the European Union FP7 IEF grant Nr. 624351.

\bibliography{3C293_ll}

\appendix

\section{X-RAY LUMINOSITIES OF MOHEGS}

Figure \ref{h2lx} shows the relation between the warm \mh luminosity and the X-ray luminosity of the host galaxy in MOHEGs, and was discussed in Section 5.3.2. Here we describe the process by which we determined the X-ray luminosity for the objects in this figure. The \mh luminosities were taken from  seven references: \citet{ogle10}, \citet{guillard12}, \citet{donahue11}, \citet{rampazzo13}, \citet{ogle12}, \citet{egami06}, and \citet{roussel07}. We retrieved the observations listed in Table \ref{xray_obs} from the Chandra archive and reprocessed them with  using \ciao version 4.5 to create  a new level 2 events file in the same manner as the 3C\,293 observation. We typically used just the longest observation, except in cases where only short observations existed. There we processed the available observations and combined the spectra. We used SDSS images (or 2MASS images where not available) to determine apertures the size of the stellar disk. The apertures were centered on the 5$-$8\,keV point central point source.  X-ray spectra were extracted using the \specextract task over the galactic disk excluding the central AGN. In the cases where we extracted spectra from multiple observations of the same galaxy, we combined them at this point. Each spectrum was fit with thermal (\apec) or power law models to determine the 0.5$-$8\,keV luminosities, given in Table \ref{xray_fits}. For the six sources with insufficient counts to fit a spectrum, we use the tool WebPIMMS\footnote{\url{http://heasarc.gsfc.nasa.gov/cgi-bin/Tools/w3pimms/w3pimms.pl}} to estimate the luminosity, assuming an APEC model with a temperature similar to that derived from fits to galaxies with similar hardness ratios: (logT=6.95 or kT=0.768\,keV for 3C\,315, 3C\,326, 3C\,424, and 3C\,459; logT=7.05, kT=0.967\,keV for 3C\,436; logT=7.65, kT=3.849\,keV for NGC\,5273). The galaxies typically fall into one of three types: radio galaxies, brightest cluster galaxies, and local early type galaxies. The exceptions are indicated on Figure \ref{h2lx}.

\clearpage
\LongTables
\setcounter{table}{2}
\begin{deluxetable}{lclrccr}
\tabletypesize{\scriptsize}
\tablecaption{Parameters of X-ray Spectral Fits\label{fits}}
\tablewidth{0pt}
\tablehead{
\colhead{Region} 	& \colhead{Model\tablenotemark{a}} 		& \multicolumn{2}{c}{Parameters} & \multicolumn{3}{c}{Luminosity\tablenotemark{d}}  \\
\cline{3-4} \cline{5-7}
\colhead{}			& \colhead{}	& \colhead{Name\tablenotemark{b}} & \colhead{Value\tablenotemark{c}} & \colhead{Component} & \colhead{Range (keV)} & \colhead{(10$^{40}$\,erg\,s$^{-1}$)} 
}
\startdata
Nucleus	& (1*) & Photon Index$_{\rm abs.}$\tablenotemark{e}				&	1.32$\pm$0.36	& Power Law 		&	0.4$-$8.0	& $874^{+128}_{-277}$~~~~\,	\\
		& 	& $S_{\rm PL, 1\,keV, abs.} {\rm (nJy)}$						&      148$^{+81}_{-77}$ & (absorbed)	&	0.4$-$2.0	& $271^{+156}_{-144}$~~~~\,	\\	
		& 	& N$_{H} $	(10$^{22}$\,cm$^{-2}$)						&	9.41$\pm$1.84	&				&	2.0$-$8.0	& $567^{+61  }_{-118}$~~~~\,	\\
\cline{5-7}\\
		& 	& Photon Index$_{\rm unabs.}$\tablenotemark{e}				&	0.57$\pm$0.87	& Power Law 		&	0.4$-$8.0	& $39.1^{+37.1}_{-16.5}$~\,	\\
		& 	& $S_{\rm PL, 1\,keV, unabs.} {\rm (nJy)}$						&      2.46$^{+1.00}_{-1.03}$ & (unabsorbed) &	0.4$-$2.0	& $4.96^{+1.87}_{-1.54}$	\\	
		& 	& 													&				&				&	2.0$-$8.0	& $36.4^{+40.0}_{-15.7}$~\,	\\
\cline{5-7}\\
		& 	& $kT{\rm (keV)}$ 										&	1.03$\pm$0.10	&  Thermal (\apec)	&	0.4$-$8.0	& $2.33^{+0.65}_{-0.69}$	\\
		& 	& N$_{\apec} {\rm (10^{-6}\,cm^{-5})}$\tablenotemark{f}			& 	2.08$\pm$0.66	&				&	0.4$-$2.0	& $2.15^{+0.60}_{-0.62}$	\\
		&	&  $\chi^2$/dof											&  	48.39/86		&				&	2.0$-$8.0	& $0.16^{+0.08}_{-0.07}$	\\
\cline{2-7}\\
		& (2)	& Photon Index\tablenotemark{e}	 						&	0.50$\pm$0.22 	& Power Law		&	0.4$-$8.0	&  $558^{+28}_{-70}$~~~~~~	\\ 
		& 	& $S_{PL, 1\,keV} {\rm (nJy)}$								&	32.81$^{+12.45}_{-12.73}$	&	&	0.4$-$2.0	&  $65.4^{+21.6}_{-23.3}$~\,	\\
		& 	& N$_{H} $	(10$^{22}$\,cm$^{-2}$)						&	4.44$\pm$0.77	&				&	2.0$-$8.0	& $486^{+25}_{-50}$	~~~~~   \\
		& 	& $\chi^2$/dof											&	165.5/90		& 					&			& 	\\
\cline{1-7} \\
NC		& (2*)	& Photon Index\tablenotemark{e}	 					&	0.77$\pm$0.24 	& Power Law		&	0.4$-$8.0	&  $739^{+48}_{-146}$~~~~\,	\\ 
		& 	& $S_{PL, 1\,keV} {\rm (nJy)}$								&	$63.6^{+24.2}_{-26.1}$	&		&	0.4$-$2.0	&  $122^{+44}_{-47}$~~~~~~	\\
		& 	& N$_{H} $	(10$^{22}$\,cm$^{-2}$)						&	6.28$\pm$0.93	&				&	2.0$-$8.0	&  $604^{+35}_{-74}$~~~~~~   \\
		& 	& $\chi^2$/dof											&	70.73/76		& 				&			& 	\\
\cline{1-7} \\
NE0		& (3*) & Photon Index\tablenotemark{e}	 						&	0.67$\pm$0.20 	& Power Law		&	0.4$-$8.0	& $34.5^{+6.9}_{-5.9}$~~\,	\\ 
		& 	& $S_{PL, 1\,keV} {\rm (nJy)}$								&	$2.61^{+0.44}_{-0.47}$	&		&	0.4$-$2.0	& $5.03^{+0.70}_{-0.85}$	\\
		& 	& $\chi^2$/dof											&	3.02/5		& 				&	2.0$-$8.0	& $29.5^{+7.3}_{-5.7}$~~\, 	\\
\cline{1-7} \\
Host: East & (3) & Photon Index\tablenotemark{e}	 						&	2.22$\pm$0.27 	& Power Law		&	0.4$-$8.0	& 7.30$^{+1.37}_{-1.11}$	\\ 
		& 	& $S_{PL, 1\,keV} {\rm (nJy)}$								&	2.33$^{+0.28}_{-0.26}$	&		&	0.4$-$2.0	& 4.60$^{+0.40}_{-0.72}$	\\
		& 	& $\chi^2$/dof											&	11.73/5		& 				&	2.0$-$8.0	& 2.78$^{+1.32}_{-0.92}$ 	\\
\cline{2-7}\\
		& (4*) & Photon Index\tablenotemark{e}							&	2.10$\pm$0.52	& Power Law		&	0.4$-$8.0	& $5.20^{+2.95}_{-1.37}$	\\
		& 	& $S_{PL, 1\,keV} {\rm (nJy)}$								&	1.47$^{+0.50}_{-0.35}$	&		&	0.4$-$2.0	& $3.01^{+0.86}_{-0.99}$	\\	
		& 	&													&				&				&	2.0$-$8.0	& $2.94^{+4.31}_{-1.14}$	\\
\cline{5-7}\\
		& 	& $kT{\rm (keV)}$ 										&	0.74$\pm$0.16	&  Thermal (\apec)	&	0.4$-$8.0	& $1.71^{+0.71}_{-0.68}$	\\
		& 	& N$_{\apec} {\rm (10^{-6}\,cm^{-5})}$\tablenotemark{f}			& 	1.29$\pm$0.50	&				&	0.4$-$2.0	& $1.65^{+0.45}_{-0.79}$	\\
		&	&  $\chi^2$/dof											&  	3.73/	3		&				&	2.0$-$8.0	& $0.050^{+0.059}_{-0.031}$	\\
\cline{2-7}\\
		& (5)	& $kT{\rm (keV)}$ 										&	0.23$\pm$0.11	&  Thermal (\apec)	&	0.4$-$8.0	& $1.48^{+1.11}_{-0.97}$	\\
		& 	& N$_{\apec} {\rm (10^{-6}\,cm^{-5})}$\tablenotemark{f}			& 	2.17$\pm$0.88	&	(0.23\,keV)	&	0.4$-$2.0	& $1.48^{+1.06}_{-0.99}$	\\
		&	& 													&  				&				&	2.0$-$8.0	& $0.0005^{+0.0043}_{-0.0005}$	\\
\cline{5-7}\\
		& 	& $kT{\rm (keV)}$ 										&	0.96$\pm$0.16	&  Thermal (\apec)	&	0.4$-$8.0	& $2.66^{+0.76}_{-0.72}$	\\
		& 	& N$_{\apec} {\rm (10^{-6}\,cm^{-5})}$\tablenotemark{f}			& 	2.23$\pm$0.65	&	(0.96\,keV)	&	0.4$-$2.0	& $2.52^{+0.79}_{-0.72}$	\\
		&	&  $\chi^2$/dof											&  	10.25/3		&				&	2.0$-$8.0	& $0.15^{+0.04}_{-0.04}$	\\
\cline{1-7} \\
Host: West & (3) & Photon Index\tablenotemark{e}	 						&	3.24$\pm$0.31 	& Power Law		&	0.4$-$8.0	& $7.84^{+0.98}_{-0.96}$	\\ 
		& 	& $S_{PL, 1\,keV} {\rm (nJy)}$								&	2.65$^{+0.35}_{-0.33}$	&		&	0.4$-$2.0	& $7.10^{+0.85}_{-0.99}$	\\
		& 	& $\chi^2$/dof											&	13.72/5		& 				&	2.0$-$8.0	& $0.85^{+0.39}_{-0.29}$ 	\\
\cline{2-7} \\
		& (4) & Photon Index\tablenotemark{e}							&	3.70$\pm$0.88	& Power Law		&	0.4$-$8.0	& $4.47^{+1.76}_{-1.65}$	\\
		& 	& $S_{PL, 1\,keV} {\rm (nJy)}$								&	1.36$^{+0.58}_{-0.62}$	&		&	0.4$-$2.0	& $4.14^{+1.43}_{-1.50}$	\\	
		& 	&													&				&				&	2.0$-$8.0	& $0.27^{+0.68}_{-0.22}$	\\
\cline{5-7} \\
		& 	& $kT{\rm (keV)}$ 										&	0.97$\pm$0.13	&  Thermal (\apec)	&	0.4$-$8.0	& $2.42^{+0.80}_{-0.89}$	\\
		& 	& N$_{\apec} {\rm (10^{-6}\,cm^{-5})}$\tablenotemark{f}			& 	2.09$\pm$0.80	&				&	0.4$-$2.0	& $2.25^{+0.73}_{-0.73}$	\\
		&	&  $\chi^2$/dof											&  	3.87/3		&				&	2.0$-$8.0	& $0.15^{+0.09}_{-0.07}$	\\
\cline{2-7} \\
		& (5*)	& $kT{\rm (keV)}$ 									&	0.42$\pm$0.12	&  Thermal (\apec)	&	0.4$-$8.0	& $2.76^{+0.71}_{-0.85}$	\\
		& 	& N$_{\apec} {\rm (10^{-6}\,cm^{-5})}$\tablenotemark{f}			& 	2.63$\pm$0.81	&	(0.42\,keV)	&	0.4$-$2.0	& $2.74^{+0.65}_{-0.68}$	\\
		&	& 													&  				&				&	2.0$-$8.0	& $0.016^{+0.017}_{-0.012}$	\\
\cline{5-7} \\
		& 	& $kT{\rm (keV)}$ 										&	1.34$\pm$0.13	&  Thermal (\apec)	&	0.4$-$8.0	& $2.58^{+0.76}_{-0.77}$	\\
		& 	& N$_{\apec} {\rm (10^{-6}\,cm^{-5})}$\tablenotemark{f}			& 	3.11$\pm$0.85	&	(1.34\,keV)	&	0.4$-$2.0	& $2.20^{+0.65}_{-0.64}$	\\
		&	&  $\chi^2$/dof											&  	1.10/3		&				&	2.0$-$8.0	& $0.38^{+0.12}_{-0.12}$	\\
\cline{1-7}  \\
Host: North+South & (3)	& Photon Index\tablenotemark{e}	 				&	2.22$\pm$0.39 	& Power Law		&	0.4$-$8.0	& $3.11^{+0.81}_{-0.54}$	\\ 
		& 	& $S_{PL, 1\,keV} {\rm (nJy)}$								&	$0.99^{+0.15}_{-0.15}$	&		&	0.4$-$2.0	& $1.91^{+0.31}_{-0.28}$	\\
		& 	& $\chi^2$/dof											&	13.52/6		& 				&	2.0$-$8.0	& $1.20^{+0.80}_{-0.52}$ 	\\
\cline{2-7}  \\
		& (4*) & Photon Index\tablenotemark{e}							&	-0.37$\pm$1.64	& Power Law		&	0.4$-$8.0	& $12.19^{+46.07}_{-9.17}$	\\
		& 	& $S_{PL, 1\,keV} {\rm (nJy)}$								&	$0.30^{+0.21}_{-0.16}$	&		&	0.4$-$2.0	& $0.72^{+0.28}_{-0.32}$	\\	
		& 	&													&				&				&	2.0$-$8.0	& $11.42^{+42.22}_{-9.35}$	\\
\cline{5-7}  \\
		& 	& $kT{\rm (keV)}$ 										&	0.39$\pm$0.08	&  Thermal (\apec)	&	0.4$-$8.0	& 1.31$^{+0.31}_{-0.32}$	\\
		& 	& N$_{\apec} {\rm (10^{-6}\,cm^{-5})}$\tablenotemark{f}			& 	1.34$\pm$0.42	&				&	0.4$-$2.0	& 1.31$^{+0.26}_{-0.32}$	\\
		&	&  $\chi^2$/dof											&  	3.63/4		&				&	2.0$-$8.0	& 0.0056$^{+0.0043}_{-0.0033}$	\\
\cline{2-7} \\
		& (6)	& $kT{\rm (keV)}$ 										&	0.57$\pm$0.09	&  Thermal (\apec)	&	0.4$-$8.0	& $1.54^{+0.24}_{-0.25}$		\\
		& 	& N$_{\apec} {\rm (10^{-6}\,cm^{-5})}$\tablenotemark{f}			& 	1.21$\pm$0.19	&				&	0.4$-$2.0	& $1.52^{+0.22}_{-0.23}$		\\
		&	&  $\chi^2$/dof											&  	12.59/6		&				&	2.0$-$8.0	& $0.022^{+0.011}_{-0.009}$	\\
\cline{1-7}   \\
NW1 & (3) & Photon Index\tablenotemark{e}								&	1.63$\pm$0.24	& Power Law		&	0.4$-$8.0	& $7.05^{+1.53}_{-1.14}$	\\ 
		& 	& $S_{PL, 1\,keV} {\rm (nJy)}$								&	1.50$^{+0.21}_{-0.22}$	&		&	0.4$-$2.0	& $2.84^{+0.42}_{-0.41}$	\\
		& 	& $\chi^2$/dof											&	3.85/4		& 				&	2.0$-$8.0	& $4.14^{+1.54}_{-1.06}$	\\
\cline{2-7}   \\
		& (6*) & $kT{\rm (keV)}$ 										&	3.58$\pm$1.24	& Thermal (\apec)	&	0.4$-$8.0	& $5.85^{+1.40}_{-1.19}$	\\
		& 	& N$_{\apec} {\rm (10^{-6}\,cm^{-5})}$\tablenotemark{f}			&	8.07$\pm$1.41	&				&	0.4$-$2.0	& $2.98^{+0.40}_{-0.38}$	\\
		&	& $\chi^2$/dof											&	1.65/4		&				&	2.0$-$8.0	& $2.94^{+1.05}_{-1.19}$	\\
\cline{1-7}   \\
NW2 & (3*) & Photon Index\tablenotemark{e}								&	2.51$\pm$0.40	& Power Law		&	0.4$-$8.0	& $4.36^{+0.88}_{-0.74}$	\\ 
		& 	& $S_{PL, 1\,keV} {\rm (nJy)}$								&	1.48$^{+0.22}_{-0.23}$	&		&	0.4$-$2.0	& $3.04^{+0.55}_{-0.50}$	\\
		& 	& $\chi^2$/dof											&	2.31/4		& 				&	2.0$-$8.0	& $1.20^{+0.89}_{-0.52}$	\\
\cline{2-7}   \\
		& (6) & $kT{\rm (keV)}$ 										&	3.26$\pm$0.92	& Thermal (\apec)	&	0.4$-$8.0	& $4.16^{+1.30}_{-1.13}$	\\
		& 	& N$_{\apec} {\rm (10^{-6}\,cm^{-5})}$\tablenotemark{f}			&	5.96$\pm$1.54	&				&	0.4$-$2.0	& $2.24^{+0.42}_{-0.43}$	\\
		&	& $\chi^2$/dof											&	6.99/4		&				&	2.0$-$8.0	& $1.97^{+0.93}_{-0.84}$	\\
\cline{1-7}   \\
SE1 & (3*) & Photon Index\tablenotemark{e}								&	1.88$\pm$0.37	& Power Law		&	0.4$-$8.0	& $5.64^{+1.87}_{-1.07}$	\\ 
		& 	& $S_{PL, 1\,keV} {\rm (nJy)}$								&	1.45$\pm$0.22	&				&	0.4$-$2.0	& $2.80^{+0.48}_{-0.47}$	\\
		& 	& $\chi^2$/dof											&	0.39/4		& 				&	2.0$-$8.0	& $2.78^{+1.84}_{-1.09}$	\\
\cline{2-7}   \\
		& (6)	& $kT{\rm (keV)}$ 										&	4.29$\pm$1.12	& Thermal (\apec)	&	0.4$-$8.0	& $5.43^{+1.32}_{-1.35}$	\\
		& 	& N$_{\apec} {\rm (10^{-6}\,cm^{-5})}$\tablenotemark{f}			&	7.29$\pm$1.43	&				&	0.4$-$2.0	& $2.52^{+0.40}_{-0.40}$	\\
		&	& $\chi^2$/dof											&	0.95/4		&				&	2.0$-$8.0	& $2.93^{+0.94}_{-0.93}$	
\enddata
\tablenotetext{a}{Each model has an additional overall absorber fixed to the MW N$_{H}=1.27\times10^{20}\,{\rm cm^{-2}}$. Model (1): \apec+absorbed power law+unabsorbed power law; Model (2): absorbed power law; Model (3): unabsorbed power law; Model (4): \apec + power law; Model (5): \apec + \apec; Model (6): \apec. When multiple models are given, the model number with * indicates the best fit. }
\tablenotetext{b}{N are component normalizations.}
\tablenotetext{c}{1 sigma errors are given.}
\tablenotetext{d}{\,Luminosity is given of the unabsorbed components.}
\tablenotetext{e}{Photon indices, $\Gamma$, are defined in the sense that $P_{E}{\rm (photons~s^{-1}\,keV^{-1})}\propto E^{-\Gamma}$ and relate to the spectral index and flux density with $F_{\nu} \propto \nu^{-\Gamma+1}\propto \nu^{-\alpha}$.}
\tablenotetext{f}{\apec normalization is given in units of 10$^{-14} \int n_{e} n_{H} dV / (4 \pi (D_A (1+z))^2)$. In these fits, z=0.045 and abundance is fixed to solar.}
\end{deluxetable}

\clearpage
\setcounter{table}{6}
\begin{deluxetable}{lllrrrllcl}
\tabletypesize{\scriptsize}
\tablecaption{\chandra Observations of MOHEGs\label{xray_obs}}
\tablewidth{0pt}
\tablehead{
\colhead{Galaxy} & \colhead{Type\tablenotemark{a}} & \colhead{Instrument} & \colhead{OID} & \colhead{Exposure} & \colhead{Date} & \multicolumn{4}{c}{Aperture} \\
\cline{7-10}
\colhead{}		   &	\colhead{}		&	\colhead{}	       &  \colhead{}	&	\colhead{(ks)}     &	\colhead{}      & \colhead{RA(J2000)}
	&	\colhead{Dec(J2000)} & \colhead{Size\tablenotemark{b}} & \colhead{AGN\tablenotemark{c}}
}
\startdata
3C\,31	& RG	&	ACIS-S	& 	 2147	& 	44.41	&	2000-11-06	& 	01:07:24.959	& 	$+$32:24:45.21		& $~\,30''\times~\,25''$(320$^{\circ}$)	&	$2\farcs5$	 \\
3C\,84	& BCG/RG &	ACIS-S	&	 4952	&	164.24	&	2004-10-14	&	03:19:48.160	&	$+$41:30:42.11		& $80''$	&	$2\farcs5$	 \\
3C\,218	& BCG/RG &	ACIS-S	& 	 4970	&	98.82	& 	2004-10-22	& 	09:18:05.688	& 	$-$12:05:43.39		& $20''$	&	$2\farcs5$	 \\
3C\,236	& RG	&	ACIS-I	&	10249	&	40.50	&	2009-01-14	&	10:06:01.735	&	$+$34:54:10.43		& $20''$	&	$2\farcs5$	 \\
		& 		&	ACIS-I	&	10246	&	29.38	&	2009-03-10	&				&					&		&			 \\
3C\,270	& RG	&	ACIS-S	&	 9569	&	100.00	&	2008-02-12	&	12:19:23.245	&	$+$05:49:29.63		& $100''$	&	$2\farcs5$	 \\
3C\,272.1	& RG	&	ACIS-S	&	 5908	& 	46.08	&	2005-05-01	&	12:25:03.707	&	$+$12:53:12.92		& $125''$	&	$2\farcs5$	 \\
3C\,293	& RG	&	ACIS-S	&	12712	& 	67.81	&	2010-11-16	&	13:52:17.821	&	$+$31:26:46.50		& $30''$	&	$2\farcs5$	 \\
3C\,305	& RG	&	ACIS-S	&	12797	& 	28.66	&	2011-01-03	&	14:49:21.625	&	$+$63:16:14.43		& $40''$	&	$2\farcs5$	 \\
3C\,310	& RG	&	ACIS-S	&	11845	& 	57.58	&	2010-04-09	&	15:04:57.179	&	$+$26:00:58.33		& $~\,9''$	&	$1\farcs0$	 \\
3C\,315	& RG	&	ACIS-S	&	 9313	& 	7.67		&	2007-12-10	&	15:13:40.096	&	$+$26:07:31.86		& $~\,5''$	&	$1\farcs0$	 \\
3C\,317	& BCG/RG &	ACIS-S	&	 5807	& 	126.95	&	2006-03-24	&	15:16:44.498	&	$+$07:01:17.62		& $35''$	&	$2\farcs5$	 \\
3C\,321	& RG	&	ACIS-S	&	 3138	&	47.13	&	2010-02-19	&	15:31:43.492	&	$+$24:04:18.96		& $18''$	&	$1\farcs5\tablenotemark{d}$ \\
3C\,326	& RG	&	ACIS-I	&	10242	& 	27.51	&	2009-05-07	&	15:52:09.140	&	$+$20:05:47.24		& $14''$	&	$1\farcs0$	 \\
		&		&	ACIS-I	&	10908	& 	18.30	&	2009-05-10	&				&					&		&			 \\
3C\,338	& BCG/RG &	ACIS-I	&	10748	& 	40.58	&	2009-11-19	&	16:28:38.202	&	$+$39:33:04.70		& $45''$	&	$2\farcs5$	 \\
3C\,386	& RG	&	ACIS-I	&	10232	& 	29.29	&	2008-11-29	&	18:38:26.251	&	$+$17:11:49.94		& $25''$	&	$1\farcs5$	 \\
3C\,405	& BCG/RG &	ACIS-I	&	 5831	& 	51.09	&	2005-02-16	&	19:59:28.296	&	$+$40:44:01.99		& $25''$	&	$2\farcs5$	 \\
3C\,424	& RG	&	ACIS-S	&	12743	& 	7.95		&	2011-04-15	&	20:48:12.099	&	$+$07:01:17.05		& $10''$	&	$1\farcs0$	 \\
3C\,433	& RG	&	ACIS-S	&	 7881	& 	37.17	&	2007-08-28	&	21:23:44.565	&	$+$25:04:27.56		& $20''$	&	$2\farcs5$	 \\
3C\,436	& RG	&	ACIS-S	& 	 9318	& 	8.04		& 	2008-01-08	& 	21:44:11.700	& 	$+$28:10:19.00		& $10''$	&	$1\farcs5$	 \\
		&		&	ACIS-S	&	12745	& 	7.95		&	2011-05-27	&				&					&		&			 \\
3C\,459	& RG	&	ACIS-S	&	12734	& 	7.95		&	2011-10-13	&	23:16:35.230	&	$+$04:05:18.50		& $10''$	&	$1\farcs0$	 \\
4C\,12.50	& RG	&	ACIS-S	&	   836	& 	25.35	&	2000-02-24	&	13:47:33.360	&	$+$12:17:24.04		& $20''$	&	$2\farcs0$	 \\
IC\,5063	& RG	&	ACIS-S	&	 7878	& 	34.10	&	2007-06-15	&	20:52:02.402	&	$-$57:04:07.58		& $100''\times~\,50''$(295$^{\circ}$)	&	$2\farcs5$	 \\
NGC\,4258 & Spiral	&	ACIS-S	&	   350	&	14.04	&	2000-04-17	&	12:18:57.518	&	$+$47:18:14.38		& $250''\times100''$(335$^{\circ}$)	&	$4\farcs0$	 \\
		&		&	ACIS-S	&	 1618	&	22.00	&	2001-05-28	&				&					&		&			 \\
\cline{1-10}
\multicolumn{10}{c}{}	\\
\multicolumn{10}{c}{\citet{donahue11}}	\\
\cline{1-10}
2A0335+096 & BCG	&	ACIS-S	&	 9792	& 	33.74	& 	2007-12-20	& 	03:38:40.861	& 	$+$09:57:57.17		& $25''$	&	$2\farcs0$	 \\
		&		&	ACIS-S	&	 7939	&	49.54	&	2007-12-16	&				&					&		&			 \\
Abell\,478	& BCG	&	ACIS-S	&	 1669	&	42.39	&	2001-01-27	&	04:13:25.300	&	$+$10:27:54.49		& $14''$	&	$1\farcs0$	 \\
Abell\,1068& BCG	&	ACIS-S	&	 1652	&	26.84	&	2001-02-04	&	10:40:44.499	&	$+$39:57:11.13		& $~\,15''\times~7\farcs5$(320$^{\circ}$)&	$1\farcs0$	 \\
Abell\,1795& BCG	&	ACIS-S	&	   493	&	19.63	&	2001-04-27	&	13:48:52.505	&	$+$26:35:34.59		& $10''$	&	$1\farcs0$	 \\
		& 		&	ACIS-S	&	   494	&	19.52	&	2001-01-12	&				&					&		&			\\
Abell\,1835& BCG	&	ACIS-I	&	 6880	& 	117.92	& 	2006-08-25	& 	14:01:02.081	& 	$+$02:52:42.34		& $~\,7''$	&	$1\farcs0$	 \\
Abell\,2597& BCG	&	ACIS-S	&	 7329	& 	60.11	& 	2006-05-04	& 	23:25:19.735	& 	$-$12:07:27.18		& $15''$	&	$1\farcs0$	 \\
MS0735.6+7421 & BCG 	& ACIS-I	&	 10470	& 	141.96	& 	2009-06-16	& 	07:41:44.346	& 	$+$74:14:39.44		& $10''$	&	$1\farcs0$	 \\
PKS\,0745-19 & BCG & ACIS-S		&	 12881	& 	118.07	& 	2011-01-27	& 	07:47:31.325	& 	$-$19:17:40.35		& $12''$	&	$1\farcs0$	 \\
\cline{1-10}
\multicolumn{10}{c}{}	\\
\multicolumn{10}{c}{\citet{rampazzo13}}	\\
\cline{1-10}
IC\,1459	& ETG	&	ACIS-S	&	2196   	& 	58.83	&	2001-08-12	&	22:57:10.607	&	$-$36:27:44.20		& $120''\times~\,90''$($275^{\circ}$)		&	$2\farcs5$	 \\
NGC\,1052 & ETG	&	ACIS-S	&	5910   	& 	59.20	&	2005-09-18	&	02:41:04.839	&	$-$08:15:20.75		& $~\,75''\times~\,50''$($300^{\circ}$)	&	$2\farcs5$	 \\
NGC\,4036 & ETG	&	ACIS-S\tablenotemark{e} & 6783 & 13.72	&	2006-07-24	&	12:01:26.753	&	$+$61:53:44.81		& R:$68''\times69''$($255^{\circ}$)		&	$1\farcs5$	 \\
NGC\,4477 & ETG	&	ACIS-S	&	9527   	& 	37.68	&	2008-04-27	&	12:30:02.198	&	$+$13:38:11.77		& $~\,90''\times~\,80''$($195^{\circ}$)	&	$1\farcs5$	 \\
NGC\,4550 & ETG	&	ACIS-S	&	8050 	& 	5.09		&	2008-02-19	&	12:35:30.622	&	$+$12:13:15.11		& $~\,70''\times~\,20''$($359^{\circ}$)	&	$1\farcs0$	 \\
		& 		&	ACIS-S	&	8058 	&	5.50		&	2008-02-18	&				&					&		&			\\
		& 		&	ACIS-S	& 	8098 	&	5.09		&	2008-02-24	&				&					&		&			\\
NGC\,4697 & ETG	&	ACIS-S	&	784   	& 	39.26	&	2000-01-15	&	12:48:35.878	&	$-$05:48:02.56		& $110''\times~\,70''$($250^{\circ}$)		&	$2\farcs0$	 \\
		& 		&	ACIS-S	&	4727   	&	39.92	&	2003-12-26	&				&					&		&			\\
		& 		&	ACIS-S	&	4728   	&	35.68	&	2004-01-06	&				&					&		&			\\
		& 		&	ACIS-S	&	4729   	&	38.10	&	2004-02-12	&				&					&		&			\\
		& 		&	ACIS-S	&	4730   	&	40.05	&	2004-08-18	&				&					&		&			\\
NGC\,5018 & ETG	&	ACIS-S	&	2070   	& 	30.89	&	2001-04-14	&	13:13:01.032	&	$-$19:31:05.65		& $~\,60''\times~\,45''$($275^{\circ}$)	&	$1\farcs5$	 \\
NGC\,5044 & ETG	&	ACIS-S	&	9399   	& 	82.68	&	2008-03-07	&	13:15:23.945	&	$-$16:23:07.33		& $72''$							&	$1\farcs5$	 \\
NGC\,5077 & ETG	&	ACIS-S	&	11780   	& 	28.68	&	2010-05-09	&	13:19:31.651	&	$-$12:39:25.21		& $~\,50''\times~\,40''$($190^{\circ}$)	&	$1\farcs0$	 \\
NGC\,5273 & ETG	&	ACIS-S	&	415   	& 	1.73		&	2000-09-03	&	13:42:08.354	&	$+$35:39:15.28		& $~\,70''\times~\,60''$($190^{\circ}$)	&	$1\farcs5$	 \\
NGC\,5353 & ETG	&	ACIS-I	&	5903   	& 	4.49		&	2005-04-10	&	13:53:26.693	&	$+$40:16:58.91		& $~\,50''\times~\,25''$($325^{\circ}$)	&	$1\farcs5$	 \\
NGC\,6868 & ETG	&	ACIS-I	&	3191   	& 	23.46	&	2002-11-01	&	20:09:54.014	&	$-$48:22:46.72		& $~\,80''\times~\,60''$($280^{\circ}$)	&	$1\farcs5$	 \\
		& 		&	ACIS-I	&	11753   	&	72.60	&	2009-08-19	&				&					&		&			\\
\cline{1-10}
\multicolumn{10}{c}{}	\\
\multicolumn{10}{c}{\citet{kaneda08}}	\\
\cline{1-10}
NGC\,708 & BCG	&	ACIS-S	&	  7921 	& 	110.67	& 2006-11-20	&	01:52:46.480		&	$+$36:09:06.60		& $~\,40''$	&	$1\farcs5$	 \\
NGC\,1395 & ETG	&	ACIS-I	&	799   	& 	27.37	& 1999-12-31	&	03:38:29.873		&	$-$23:01:38.90		& $75''$	&	$1\farcs5$	 \\
NGC\,1549 & ETG	&	ACIS-S	&     2077	  	& 	36.53	& 2000-11-08	&	04:15:45.175		&	$-$55:35:32.37		&$~\,80''\times~\,90''$(270$^{\circ}$)		&	$1\farcs5$	 \\
NGC\,3557 & ETG	&	ACIS-I	&	3217   	& 	37.49	& 2002-11-28	&	11:09:57.642		&	$-$37:32:20.96		& $~\,70''\times~\,50''$(210$^{\circ}$)	&	$1\farcs5$	 \\
NGC\,3894 & ETG	&	ACIS-S	&	   10389	& 	38.54	& 2009-07-20	&	11:48:50.306		&	$+$59:24:56.38		& $~\,60''\times~\,40''$(200$^{\circ}$)	&	$1\farcs5$	 \\
NGC\,4696 & BCG	&	ACIS-S	&	 2931  	& 	9.63		& 2002-09-21	&	21:18:33.025		&	$+$26:26:49.03		& $~\,60''\times~\,40''$(245$^{\circ}$)	&	$2\farcs0$	 \\
NGC\,4589 & ETG	&	ACIS-S\tablenotemark{e}	& 6785 & 13.77	 & 2006-08-31	&	12:37:24.871		&	$+$74:11:30.92		& R:$110''\times65''$(300$^{\circ}$)	&	$1\farcs0$	 \\
NGC\,7052& ETG	&	ACIS-S	&	4954   	& 	89.05	& 2004-04-01	&	12:48:49.250		&	$-$41:18:39.00		& $~\,45''\times~\,35''$(275$^{\circ}$)	&	$2\farcs0$	 \\
\cline{1-10}
\multicolumn{10}{c}{}	\\
\multicolumn{10}{c}{}	\\
\multicolumn{10}{c}{\citet{ogle12}, \citet{egami06}, \citet{alatalo11}}	\\
\cline{1-10}
PKS\,1138-26 & PC	&	ACIS-S	&	898		&	39.47	& 2000-06-06	& 11:40:48.353			& $-$26:29:08.52	&	$~\,10''$ 	&$1\farcs0$	\\
Z3146 	      & BCG &	ACIS-I	&	909		&	46.01	& 2000-05-10	& 10:23:39.600			& $+$04:11:10.00 	&      $~\,10''$ 	&$2\farcs0$	\\ 
NGC\,1266     & ETG &	ACIS-S	&	11578	&	29.65	& 2009-09-20 	& 03:16:00.769			& $-$02:25:38.54 	& $~\,40''\times~\,30''$(108$^{\circ}$) &$1\farcs5$	 
\enddata
\tablenotetext{a}{RG: Radio Galaxy; BCG: Brightest Cluster Galaxy; ETG: Early Type Galaxy; PC: Protocluster}
\tablenotetext{b}{Apertures were either circular, in which case radii are given, or ellipses, in which case size is given as semi-major axis$\times$semi-minor axis (position angle given counterclockwise from North). In two cases, NGC4589 and NGC4036, we had to use a rectangle (R).}
\tablenotetext{c}{Radius of the exclusion region around the AGN.}
\tablenotetext{d}{The nucleus of both 3C\,321 and its companion were excluded.}
\tablenotetext{e}{Only subarray mode is available, so the extraction aperture is the rectangle overlap region between the ellipse derived from the stellar emission and the subarray.}
\end{deluxetable}

\clearpage
\begin{deluxetable}{lrrcccrclc}
\tabletypesize{\scriptsize}
\tablecaption{MOHEGs Parameters\label{xray_fits}}
\tablewidth{0pt}
\tablehead{
\colhead{Galaxy} & \colhead{D\tablenotemark{a}} & \colhead{Log(L$_{\mhb}$}	&	\colhead{Ref.\tablenotemark{c}} &\colhead{Log(L$_{\rm AGN}$} &	\colhead{Ref.\tablenotemark{c}} & \multicolumn{4}{c}{Host X-ray Emission} \\
\cline{7-10}
\colhead{}		 & \colhead{(Mpc)}	 			& \colhead{(erg/s))\tablenotemark{b}}	&\colhead{}		& \colhead{(erg/s))}			&\colhead{}					&  \colhead{Counts} & \colhead{Model\tablenotemark{d}}		   &\colhead{Model }	&	 \colhead{Log(L$_{\rm X}$}	\\
\colhead{}		 & \colhead{}					 & \colhead{}						& \colhead{}		& \colhead{(2-10\,keV)}		& \colhead{}					 & \colhead{(0.5-8\,kev)} &	 \colhead{}						& \colhead{Parameters} 	& 	\colhead{(erg/s))\tablenotemark{e}}	
}
\startdata
3C\,31	&	  74~~&	40.32~~	& (1)		& 40.67	& (1)			& 	1754	 	 &	M1	&	kT=0.91\,keV		& 41.07 \\
		&	  	&			& 		& 		& 			& 		 	 &		&	$\Gamma=1.94$	& 	      \\
\cline{1-10}
3C\,84	&	  76~~&	41.75~~	& (1)		& 42.91	& (1)			& 	3434419	 &	M2	&	kT1=0.93\,keV		& 44.10 \\
		&	  	&			& 		& 		& 			& 		 	 &		&	kT2=3.93\,keV		& 	      \\
\cline{1-10}
3C\,218	&	245~~&	41.10~~	& (1)		& 41.69	& (1)			& 	99092	 &	M2	&	kT1=0.78\,keV		& 43.67 \\
		&	  	&			& 		& 		& 			& 		 	 &		&	kT2=3.69\,keV	 	&   	      \\
\cline{1-10}
3C\,236	&	462~~&	41.76~~	& (2)		& 43.02	& (2)			& 	102	 	 &	M3	&	$\Gamma=0.21$	& 42.11 \\
\cline{1-10}
3C\,270	& 32$^{*}$\,&	39.26~~	& (1)		& 41.08	& (1)			& 	14218	 &	M2	&	kT1=0.78\,keV		& 40.96 \\
		&	  	&			& 		& 		& 			& 		 	 &		&	kT2=5.72\,keV		& 	      \\
\cline{1-10}
3C\,272.1	&18$^{*}$\,&	38.86~~	& (1)		& 39.34	& (1)			& 	16732	 &	M2	&	kT1=0.74\,keV		& 41.46 \\
		&	  	&			& 		& 		& 			& 		 	 &		&	kT2=6.84\,keV 		& 	      \\
\cline{1-10}
3C\,293	&	199~~&	41.76~~	& (1)		& 42.78	& (1)			& 	634	 	 &	M2	&	kT=0.39\,keV		& 41.39 \\
		&	  	&			& 		& 		& 			& 		 	 &		&	kT2=4.94\,keV	& 	      \\
\cline{1-10}
3C\,305	&	185~~&	$<$41.48~~ & (2)	& 41.23	& (2)			& 	373		 &	M5	&	kT=0.98\,keV		& 41.29 \\
\cline{1-10}
3C\,310	&	240~~&	40.85~~	& (1)		& 40.11	& (10)		& 	580		 &	M4	&	kT=1.21\,keV		& 41.47 \\
		&	  	&			& 		& 		& 			& 		 	 &		&	 Z=0.27			& 	      \\
\cline{1-10}
3C\,315	&	501~~&	41.76~~	& (1)		& 41.68	& (1)			& 	11\tablenotemark{f}	&	&				& 41.47 \\
\cline{1-10}
3C\,317	&	152~~&	40.53~~	& (1)		& 41.30	& (1)			& 	159379	 &	M2	&	kT1=1.24\,keV		& 43.30 \\
		&	  	&			& 		& 		& 			& 		 	 &		&	 kT2=2.91\,keV		& 	      \\
\cline{1-10}
3C\,321	&	441~~&	42.13~~	& (1)		& 42.08	& (1)			& 	305		 &	M1	&	kT=0.21\,keV		& 41.81  \\
		&	  	&			& 		& 		& 			& 		 	 &		&	$\Gamma=2.95$	& 	      \\
\cline{1-10}
3C\,326	&	409~~&	41.73~~	& (1)		& 40.63	& (1)			& 	48\tablenotemark{f}	&	&				& 41.37  \\
\cline{1-10}
3C\,338	&	133~~&	40.54~~	& (1)		& 40.30 	& (1)			& 	53675	 &	M4	&	kT=3.48\,keV		& 43.45 \\
		&	  	&			& 		& 		& 			& 		 	 &		&	Z=0.84			& 	      \\
\cline{1-10}
3C\,386	&	  73~~&	39.87~~	& (1)		& 39.75	& (1)			& 	97		 &	M5	&	kT=1.53\,keV		& 40.24 \\
\cline{1-10}
3C\,405	&	251~~&	41.75~~	& (1)		& 44.28	& (1)			& 	49694	 &	M5	&	kT=4.76\,keV		& 44.06 \\
\cline{1-10}
3C\,424	&	595~~&	41.97~~	& (1)		& 42.44	& (11)		& 	34\tablenotemark{f}	&	&				& 42.12 \\
\cline{1-10}
3C\,433	&	468~~&	42.13~~	& (1)		& 43.90	& (1)			& 	301		 &	M1	&	kT=0.93\,keV		& 42.62 \\
		&	  	&			& 		& 		& 			& 		 	 &		&	$\Gamma=0.37$	& 	      \\
\cline{1-10}
3C\,436	&	1000~~&	42.31~~	& (1)		& 43.53	& (1)			& 	36\tablenotemark{g}	&	&				& 42.30 \\
\cline{1-10}
3C\,459	&	1090~~&	$<$42.35~~ & (2)	& 43.24	& (2)			& 	46\tablenotemark{f}	&	&				& 42.80 \\
\cline{1-10}
4C\,12.50	&	 577~~&	42.54~~	& (2)		& 43.34	& (2)			& 	160		 &	M1	&	kT=0.75\,keV		& 42.29 \\
		&	  	&			& 		& 		& 			& 		 	 &		&	$\Gamma=0.54$	& 	      \\
\cline{1-10}
IC\,5063	&	  49~~&	$<$40.88~~ & (2)	& 42.97	& (2)			& 	1085		 &	M1	&	kT=0.71\,keV		& 41.34 \\
		&	  	&			& 		& 		& 			& 		 	 &		&	$\Gamma=-1.08$	& 	      \\
\cline{1-10}
NGC\,4258&7.2$^{\#}$&	40.59~~	& (3)		& 40.86	& (3)			& 	23365	 &	M6	&	kT1=0.24\,keV		& 40.10 \\
		&	  	&			& 		& 		& 			& 		 	 &		&	kT2=0.74\,keV		& 	      \\
		&	  	&			& 		& 		& 			& 		 	 &		&	Z2=0.19			& 	      \\
\cline{1-10}
\hline \hline 
2A0335+096 &	  153~~&	41.51$^{*}$\,& (4)	& $<$40.81~~ & (12)		& 	127023	 &	M7	&	kT1=2.49\,keV		&  43.51  \\
		&	  	&			& 		& 		& 			& 		 	 &		&	kT2=1.03\,keV		& 	      \\
		&	  	&			& 		& 		& 			& 		 	 &		&	N$_{\rm H, kT2}$=4.2${\rm E}$21\,cm$^{-2}$	& 	      \\
\cline{1-10}
Abell\,478 &	  392~~&	42.07$^{*}$\,& (4)	& $<$41.77~~ & 	(12)	& 	33502	 &	M8	&	kT=4.15\,keV		&  44.10 \\
		&	  	&			& 		& 		& 			& 		 	 &		&	N$_{\rm H}$=1.9${\rm E}$21\,cm$^{-2}$		& 	      \\
\cline{1-10}
Abell\,1068 &	  654~~&$<$42.55$^{*}$\,& (4)	& $<$43.43~~ & 	(13)	& 	10794	 &	M2	&	kT1=1.37\,keV		&   44.04  \\
		&	  	&			& 		& 		& 			& 		 	 &		&	kT2=3.65\,keV		& 	        \\
\cline{1-10}
Abell\,1795 &	  284~~&	41.74$^{*}$\,& (4)	& $<$41.24~~ & 	(12)	& 	15521	 &	M2	&	kT1=4.22\,keV		&   43.33  \\
		&	  	&			& 		& 		& 			& 		 	 &		&	kT2=1.04\,keV		&   	        \\
\cline{1-10}
Abell\,1835 &	  1272~~&$<$42.80$^{*}$\,& (4)& $<$42.78~~ & 	(12)	& 	25290	 &	M5	&	kT=4.63\,keV		&   44.64   \\
\cline{1-10}
Abell\,2597&	  373~~&	42.39$^{*}$\,& (4)	& $<$41.58~~ & 	(12)	& 	40046	 &	M6	&	kT1=1.06\,keV		&   43.83    \\
		&	  	&			& 		& 		& 			& 		 	 &		&	kT2=3.06\,keV		&    		  \\
		&	  	&			& 		& 		& 			& 		 	 &		&	Z2=0.49			& 	         \\
\cline{1-10}
MS0735.6+7421 & 1068~~&42.05$^{*}$\,& (4)	& $<$42.11~~ & 	(12)	& 	11505	 &	M5	&	kT1=3.85\,keV		&   44.07    \\
\cline{1-10}
PKS\,0745-19 	& 474~~~&42.30$^{*}$\,& (4)	& 42.11	& 	(12)		& 	80110	 &	M1	&	kT1=4.01\,keV		&   44.38   \\
		&	  	&			& 		& 		& 			& 		 	 &		&	$\Gamma=1.98$	& 	      \\
\cline{1-10}
\hline \hline 
IC\,1459	&29$^{*}$\,&	39.25~~	& (5)		& 40.87	& (14)		& 	6363	 	 &	M1	&	kT1=0.59\,keV		& 	 40.73     \\
		&	  	&			& 		& 		& 			& 		 	 &		&	$\Gamma=2.09$	& 	      \\
\cline{1-10}
NGC\,1052	&19$^{*}$\,&	39.83$^{*}$\,& (5)	& 41.20	& (14)		& 	2661	 	&	M1	&	kT1=0.61\,keV		& 40.12     \\
		&	  	&			& 		& 		& 			& 		 	&		&	$\Gamma=1.73$	& 	      \\
\cline{1-10}
NGC\,4036&24$^{+}$&	39.97$^{||}$& (5)	& 39.08	& (14)		& 	197	 	&	M2	&	kT1=0.78\,keV		& 39.62     \\
		&	  	&			& 		& 		& 			& 		  	&		&	kT2=0.11\,keV	& 	      \\
\cline{1-10}
NGC\,4477&22$^{+}$&	39.35$^{*}$\,& (5)	& 39.81	& (15)$^{*}$	& 	2060	 	&	M2	&	kT1=0.25\,keV		& 40.12     \\
		&	  	&			& 		& 		& 			& 		 	&		&	kT2=0.68\,keV		& 	      \\
\cline{1-10}
NGC\,4550&16$^{*}$\,&	38.86$^{||}$& (5)	& $<$38.36~~  & (14)	& 	92	 	&	M1	&	kT1=0.31\,keV		& 38.57     \\
		&	  	&			& 		& 		& 			& 		 	&		&	$\Gamma=0.89$	& 	      \\
\cline{1-10}
NGC\,4697&12$^{*}$\,&	38.40~~	& (5)		& 38.40	& (14)		& 	13670	 &	M1	&	kT1=0.32\,keV		& 39.85     \\
		&	  	&			& 		& 		& 			& 		 	 &		&	$\Gamma=1.73$	& 	      \\
\cline{1-10}
NGC\,5018&45$^{+}$&	39.66~~	& (5)		& $<$39.52~~  & (14)	& 	822	 	&	M1	&	kT1=0.56\,keV		& 40.42     \\
		&	  	&			& 		& 		& 			& 		 	&		&	$\Gamma=2.35$	& 	      \\
\cline{1-10}
NGC\,5044&31$^{*}$\,&	39.72~~	& (5)		&  39.44	& (14)		& 	169214	 &	M2	&	kT1=2.05\,keV		& 40.38     \\
		&	  	&			& 		& 		& 			& 		 	 &		&	kT2=0.91\,keV	& 	      \\
\cline{1-10}
NGC\,5077&40$^{+}$&	39.76~~	& (5)		& 39.73	& (16)		& 	710	 	&	M1	&	kT1=0.80\,keV		& 42.03     \\
		&	  	&			& 		& 		& 			& 		 	&		&	$\Gamma=2.37$	& 	      \\
\cline{1-10}
NGC\,5273&17$^{*}$\,&	38.88$^{\#}$& (5)	& 40.55	& (14)		& 	35\tablenotemark{h}	& &					& 39.73     \\
\cline{1-10}
NGC\,5353&36$^{+}$\,&	38.96~~	& (5)		&  $<$39.78~~	& 	i	& 	209		 &	M2	&	kT1=5.09\,keV		& 40.69     \\
		&	  	&			& 		& 		& 			& 		 	 &		&	kT2=0.75\,keV		& 	      \\
\cline{1-10}
NGC\,6868&27$^{*}$\,&	39.65$^{*}$\,& (5)	& $<$39.98~~	& (17)	& 	4357	 	&	M1	&	kT1=0.75\,keV		& 40.53     \\
		&	  	&			& 		& 		& 			& 		 	 &		&	$\Gamma=1.96$	& 	      \\
\cline{1-10}
\hline \hline 
 NGC\,708 	& 70~~	&	40.47$^{+}$& (6)	& $<$39.02~~ & (12)		& 	68804	 &	M6	&	kT1=1.06\,keV		& 42.26     \\
		&	  	&			& 		& 		& 			& 		 	 &		&	kT2=2.02\,keV		& 	      \\
		&	  	&			& 		& 		& 			& 		 	 &		&	Z2=1.73			& 	         \\
\cline{1-10}
 NGC\,1395&24$^{*}$\,&	38.92$^{+}$& (6)	& 39.06	& (14)		& 	2252	 	&	M1	&	kT1=0.84\,keV		& 40.60     \\
		&	  	&			& 		& 		& 			& 		 	 &		&	$\Gamma=1.82$	& 	      \\
\cline{1-10}
 NGC\,1549&20$^{*}$\,&	38.20$^{+}$& (6)	& 38.46	& (14)		& 	1055	 	&	M3	&	$\Gamma=1.42$	& 40.51     \\
\cline{1-10}
 NGC\,3557&45~~	&	$<$39.20$^{+}$& (6)	& 40.22	& (14)		& 	433	 	&	M1	&	kT1=0.74\,keV		&  40.53    \\
		&	  	&			& 		& 		& 			& 		 	 &		&	$\Gamma=1.91$	& 	      \\
\cline{1-10}
 NGC\,3894&45~~	&	39.93$^{+}$& (6)	& 40.72	& (18)		& 	706	 	&	M1	&	kT1=0.33\,keV		&  40.40    \\
		&	  	&			& 		& 		& 	 		& 		 	 &		&	$\Gamma=1.81$	& 	      \\
\cline{1-10}
 NGC\,4589&22$^{*}$\,&	39.36$^{+}$& (6)	& 38.90	& (14)		& 	257	 	&	M1	&	kT1=0.71\,keV		&  39.72    \\
		&	  	&			& 		& 		& 			& 		 	 &		&	$\Gamma=2.30$	& 	      \\
\cline{1-10}
 NGC\,4696&36$^{*}$\,&	39.68$^{+}$& (6)	& 40.00	& (14)		& 	386482	 &	M6	&	kT1=1.08\,keV		&  42.58    \\
		&	  	&			& 		& 		& 			& 		 	 &		&	kT2=2.53\,keV		& 	      \\
		&	  	&			& 		& 		& 			& 		 	 &		&	Z2=2.79			& 	         \\
\cline{1-10}	
 NGC\,7052&68~~	&	39.71$^{+}$& (6)	& $<$40.29~~ & (14)		& 	690	 	 &	M1	&	kT1=0.64\,keV		&  41.20    \\
		&	  	&			& 		& 		& 			& 		 	 &		&	$\Gamma=2.55$	& 	      \\
\cline{1-10}
\hline \hline 
PKS\,1138-26&17346~~ &44.92$^{||}$ & (7)	& 45.60	& 	(19)		& 	164		 &	M3	&	$\Gamma=1.10$	& 45.14  \\
\cline{1-10}
Z3146	& 1510~~	   &	43.20$^{*}$\, & (8)	& 42.91	& 	(1)		& 	10749	 &	M5	&	kT=4.71\,keV 		&  44.18 \\
\cline{1-10}	
NGC\,1266& 29.9$^{||}$ &	 40.71~~	    & (9)	& 39.53	& 	j		& 	711	 	 &	M9	&	kT=0.78\,keV		&  40.61 \\
 		&	  	&			& 		& 		& 			& 		 	 &		&	$\Gamma=4.24$	& 	      \\
		&	  	&			& 		& 		& 			& 		 	 &		&	N$_{\rm H}=4.1{\rm E}$21\,cm$^{-2}$	& 	      
 \enddata
\tablenotetext{a}{Distances are calculated assuming $H_{0}=70\,{\rm km\,s^{-1}\,Mpc^{-1}}$, $\Omega_{M}=0.3$, and $\Omega_{\Lambda}=0.7$ if z$\ge0.01$. Otherwise, distances were taken from \citet{tonry01} ($^{*}$), \citet{theureau07} ($^{+}$), \citet{ogle14} ($^{\#}$), and \citet{alatalo11} ($^{||}$).}
\tablenotetext{b}{0-0 S(0)-S(3) luminosities. When only a subset of the lines are available, we estimate L$_{\rm S(0)-S(3)}$=1.23$\times$L$_{\rm S(1)-S(3)}$ ($^{*}$), 
L$_{\rm S(0)-S(3)}$=1.55$\times$L$_{\rm S(1)+S(3)}$ ($^{+}$), L$_{\rm S(0)-S(3)}$=3.14$\times$L$_{\rm S(1)}$ ($^{\#}$), and L$_{\rm S(0)-S(3)}$=3.76$\times$L$_{\rm S(3)}$ ($^{||}$). The conversion factors were derived based on the \citet{ogle10} line luminosities.}
\tablenotetext{c}{References: (1) \citet{ogle10}, (2) \citet{guillard12}, (3) \citet{ogle14}, (4) \citet{donahue11},  (5) \citet{rampazzo13}, (6) \citet{kaneda08}, (7) \citet{ogle12}, (8) \citet{egami06}, (9) \citet{roussel07}, (10) \citet{kraft12}, (11) \citet{massaro12}, (12) \citet{russell13}, (13) \citet{wise04}, (14) \citet{pellegrini10}, (15) \citet{cappi06}, (16) \citet{gultekin12},  (17) \citet{fukazawa06}, (18) \citet{ueda05}, (19) \citet{carilli02}, and \citet{alatalo11}. $^{*}$ indicates the 2-10 keV measurement of the AGN luminosity was made with \emph{XMM-Newton} data, while $^{+}$ indicates it was made with the \emph{ASCA} satellite.} 
\tablenotetext{d}{Models used are: M1: sum of an \apec thermal model and a power law; M2: sum of 2 \apec thermal models; M3: a power law; M4: an \apec thermal model with free abundance; M5: an \apec model with fixed solar abundance; M6: the sum of two \apec model, whose hotter component has a free abundance; M7: the sum of an absorbed \apec thermal model and an unabsorbed \apec thermal model; M8 an absorbed \apec thermal model; and M9: the sum of an absorbed power law and an unabsorbed \apec thermal model.}
\tablenotetext{e}{0.5-8\,keV luminosities from spectral fits where possible.}
\tablenotetext{f}{There are insufficient counts to fit a spectrum, so we use WebPIMMS to estimate
the luminosity assuming an APEC model with logT=6.95 (0.768\,keV).}
\tablenotetext{g}{There are insufficient counts to fit a spectrum, so we use WebPIMMS to estimate
the luminosity assuming an APEC model with logT=7.05 (0.967\,keV).}
\tablenotetext{h}{There are insufficient counts to fit a spectrum, so we use WebPIMMS to estimate
the luminosity assuming an APEC model with logT=7.65 (3.849\,keV).}
\tablenotetext{i}{No literature estimate of the AGN luminosity exists and the 2-10 keV detection in an 8$''$ aperture is not significant. This limit is estimated based on a 3$\sigma$ count rate upper limit and a $\Gamma=2$ power-law model is WebPIMMS.}
\tablenotetext{j}{No literature estimate of the AGN luminosity exists. This measurement is made in a 3$''$ aperture centered on the hard source, but is likely an underestimate since there are indications that this AGN may be Compton-thick \citep{nyland13}.}
\end{deluxetable}

\end{document}